\documentstyle[12pt,mssymb,equations]{article}
\newcounter{eqnpart}  

\renewcommand{\Re}{\mbox{Re}} 
\renewcommand{\Im}{\mbox{Im}} 
\renewcommand{\cite}[1]{\mbox{$^{#1}$}} 

\newcommand{\bold}[1]{\mbox{\boldmath $#1$}}

\newcommand{\tensor}[2]  
  {(\raisebox{-.6ex}{$\stackrel{#1}{\scriptstyle #2}$})-tensor}
\newcommand{\pseudotensor}[2]  
  {(\raisebox{-.6ex}{$\stackrel{#1}{\scriptstyle #2}$})-pseudotensor}

\newcommand{\dom}{\mbox{dom}}
\newcommand{\E}{{\cal E}} 

\newcommand{\DIV}{\mbox{D$_{IV}$}}
\newcommand{\G}{{\cal G}}
\newcommand{\SP}{F}
\newcommand{\e}{\E}
\newcommand{\CW}{CW}
\newcommand{\CGPWP}{CPGWP}
\newcommand{\izz}[1]{\mbox{\footnotesize $#1$}}  
\newcommand{\Izz}[1]{\mbox{\normalsize $#1$}} 
\begin{document}
%
%
\title{Non-Analytic Extension of the Kinnersley-Chitre Group for Colliding
Plane Gravitational Waves. I}
\author{Isidore Hauser\thanks{Home address: 4500 19th Street, \#342,
	Boulder, CO 80304.}
	and
	Frederick J.\ Ernst \\
	FJE Enterprises\thanks{A private company supporting
	research of F.\ J.\ Ernst and I.\ Hauser
	in classical general relativity theory.} \\
	Rt.\ 1, Box 246A, Potsdam, NY 13676}
\date{December 28, 1992}
\maketitle
\begin{abstract}
A program is outlined concerning the set $\Sigma_{\e}$ of all solutions
of the hyperbolic Ernst equation on a $2$-dimensional manifold whose
underlying topological space $\{ (r,s) : -1 \le r < s \le 1 \}$ is the
same as the domain of all Ernst potentials $\bold{\E}$ for colliding
plane gravitational wave pairs (\CGPWP's).  The aim of the program
is to construct and apply a non-trivial extension $\bar{K}$ of the group
of Kinnersley-Chitre transformations (i.e., the representation of the
Geroch group which is due to W.\ Kinnersley and D.\ M.\ Chitre) such that
$\bar{K}$ is transitive on $\Sigma_{\e}$.  This is to be done by
employing the formalism of a homogeneous Hilbert problem which generalizes
one previously used to effect KC transformations [I.\ Hauser and F.\ J.\
Ernst, J.\ Math.\ Phys.\ {\bf 21}, 1126 (1980); F.\ J.\ Ernst, A.\
Garc\'{\i}a-D\'{\i}az and I.\ Hauser, J.\ Math.\ Phys.\ {\bf 29}, 681
(1988)].  In this first paper of a series, the aforementioned program
is completely carried out for the set $\Sigma_{\psi}$ of all real members
$\bold{\E} = - \exp(2\bold{\psi})$ of $\Sigma_{\e}$ and for an abelian
subgroup of $\bar{K}$ such that this subgroup is transitive on
$\Sigma_{\psi}$.  A simple Hilbert problem is used to obtain a new form
of the general solution for $\bold{\psi}(r,s)$ and this gneral solution
provides an independent derivation of the known result [U.\ Yurtsever,
Phys.\ Rev.\ D {\bf 38}, 1706 (1987)] for the asymptotic form of
$\bold{\psi}(r,s)$ as one approaches the axis $\{ (r,s) : -1 < r = s
< 1 \}$.  The necessary and sufficient condition that $\bold{\psi}(r,s)$
have a continuous extension to the axis is $\sqrt{1-\sigma} g_{3}(\sigma)
= \sqrt{1+\sigma} g_{2}(\sigma) \; (-1 < \sigma < 1)$ where $g_{3}(\sigma)$
and $g_{2}(\sigma)$ are those Abel transforms of the initial value
function derivatives which were introduced by I.\ Hauser and F.\ J.\
Ernst, J.\ Math.\ Phys.\ {\bf 30}, 872 (1989); 2322 (1989).
\end{abstract}

%
\newpage

\section{Introduction}
This is the first of a series of papers on a non-trivial extension
of the transformation group which was discovered by W.\ Kinnersley
and D.\ M.\ Chitre\cite{1,2,3} (KC) and which is an effective realization of
the symmetry group originally conceived by R.\ Geroch.\cite{4,5}  The
generalization we have in mind seems to be appropriate when the
field equations are of hyperbolic type, as they are, for example,
in the case of colliding plane gravitational waves.

\subsection{Basic Concepts}
The Ernst potentials\cite{6} which represent the regions of interaction of
colliding plane gravitational wave pairs (\CGPWP's) are defined
on certain two dimensional differential manifolds.\cite{7,8}  The underlying
topological space of all of these manifolds is the $R^2$-subspace
$\{ (r,s) : -1 \le r < s \le 1 \}$, and the charts which are
usually employed are the familiar mappings
$$(r,s) \rightarrow \left( r(u) , s(v) \right)$$
into the null coordinate $(u,v)$-plane.

We shall be considering the set $\Sigma_{\e}$ of all Ernst potentials\cite{9}
$\bold{\E}$ which are defined on the same manifolds as those described
above but which are not required to satisfy the colliding wave
conditions.\cite{10}  The set $\Sigma_{\e}$ includes all Ernst potentials
which represent the regions of interaction of \CGPWP's but it also
includes Ernst potentials (such as those for the Kasner metrics)
which do not represent any \CGPWP's.\cite{11}

It is the group of all KC transformations which induce {\em
permutations} of $\Sigma_{\e}$, i.e., which induce one-to-one mappings
of $\Sigma_{\e}$ onto $\Sigma_{\e}$,\cite{12} that we shall be extending.
We caution that each element of this KC group is not itself a
permutation of $\Sigma_{\e}$.  It is a permutation of a certain set
$\Sigma_{\SP}$ which is closely related to $\Sigma_{\e}$.  The KC
group will accordingly be denoted by $K(\Sigma_{\SP})$.  The set
$\Sigma_{\SP}$ is a special gauge of the $2 \times 2$ matrix spectral
potentials which were originally introduced by KC.\cite{2,13,14,15}  The
definition of this special gauge is still being studied and will
be covered in a later paper of the series.  At present, it suffices
to note that the domain of any member $\bold{F}$ of $\Sigma_{\SP}$
will contain the set of all quadruples $(r,s,\tau,z_{0})$ such that
$r,s,z_{0}$ are real and $-1 \le r < z_{0} < s \le 1$, while $\tau$
(the so-called spectral parameter) is any point in the extended
complex plane $C$ minus the real axis interval $[r,s]$.  Most previous
papers on KC transformations employ the parameter $t=(2\tau)^{-1}$ in
place of $\tau$ and no previous paper employs the parameter $z_{0}$.

To enable us to present our key concepts and methods before having to
cope with intricate mathematical problems stemming from the
nonlinearity of the Ernst equation, we shall focus attention in this
paper on the real members of $\Sigma_{\e}$, i.e., on the Ernst
potentials of vacuum Weyl metrics.  Complex Ernst potentials will be
covered in our next paper.

For the real members of $\Sigma_{\e}$ which are the subject of the
current paper, a factorization of $\bold{F}$ due to KC\cite{2} yields
\begin{equation}
\bold{F}(r,s,\tau,z_{0}) = e^{-\sigma_{3} \bold{\psi}(r,s)}
\bold{F}^{MS}(r,s,\tau) e^{-\sigma_{3} \bold{\xi}(r,s,\tau,z_{0})}
\end{equation}
where $\sigma_{3}$ is the usual Pauli spin matrix, $\bold{\psi}$ is
half of the natural logarithm of $-\bold{\E}$, the second factor
$\bold{F}^{MS}$ is the member of $\Sigma_{\SP}$ corresponding to the
Minkowski space Ernst potential $\bold{\E}^{MS} = -1$ and $\bold{\xi}$ is
a complex-valued spectral potential.  The above Eq.\ (1.1) will not
be used in this paper but we shall define, thoroughly discuss and
use $\bold{\xi}$ in Secs.\ II, III and IV.

The spectral potential $\bold{F}$ in the gauge $\Sigma_{\SP}$ is
uniquely determined by the choice of $\bold{\E} \in \Sigma_{\e}$ except
perhaps for the gauge transformations $\bold{F} \rightarrow
\bold{F} U$ where $U$ is any $2 \times 2$ unimodular matrix which
is dependent only on $\tau$ and $z_{0}$, is a holomorphic function
of $\tau$ throughout $C-\{z_{0}\}$ with no essential singularity,
is real for real $\tau$ and is the unit matrix $I$ at $\tau=\infty$.
Conversely, the Ernst potential $\bold{\E} \in \Sigma_{\e}$ corresponding
to a given $\bold{F} \in \Sigma_{\SP}$ is uniquely determined from
the second term of the following expansion in a neighborhood of
$\tau = \infty$:
\begin{equation}
\bold{F}(r,s,\tau,z_{0}) = \left( \begin{array}{cc}
0 & i \\
-i & 0
\end{array} \right)
+ (2\tau)^{-1} [\bold{H}(r,s) + i \bold{B}(z_{0})] + O(\tau^{-2})
\end{equation}
where $\bold{B}(z_{0})$ is a real symmetric matrix and $\bold{H}(r,s)$
is a matrix generalization of the Ernst potential such that (with our
conventions) $\bold{H}(-1,1)=-I$ and $\bold{\E}(r,s)$ is the lower
diagonal element of $\bold{H}(r,s)$.  There is thus an obvious
natural homomorphism of $K(\Sigma_{\SP})$ onto a group $K(\Sigma_{\e})$
of permutations of the Ernst potential set $\Sigma_{\e}$.  It is
worth recalling that the transformations $\bold{\E}^{(0)} \rightarrow
\bold{\E}$ resulting from almost every permutation in $K(\Sigma_{\e})$
are non-local and the direct calculation of $\bold{\E}$ in terms of
$\bold{\E}^{(0)}$ involves an unwieldy sequence (generally infinite)
of iterated integrations over $r$ and $s$.  In contrast, every
transformation $\bold{F}^{(0)} \rightarrow \bold{F}$ resulting from
a permutation in $K(\Sigma_{\SP})$ is local with respect to $r,s,z_{0}$
and the calculation of $\bold{F}$ requires that one solve a linear
integral equation of the Cauchy type in the $\tau$-plane.\cite{16}  The kernel
of this integral equation is a quadratic expression in $\bold{F}^{(0)}$.

Our extension of $K(\Sigma_{\SP})$ will also be a group of permutations
of $\Sigma_{\SP}$ and will be denoted by $\bar{K}(\Sigma_{\SP})$.
The corresponding extension of $K(\Sigma_{\e})$ will be denoted by
$\bar{K}(\Sigma_{\e})$.  One way of defining $\bar{K}(\Sigma_{\SP})$
involves two constructs which generalize the formalism\cite{14} previously
used by the authors for KC transformations and which will be motivated and
detailed in later papers of the series.  Brief descriptions of these
two constructs follow in (i) and (ii) below:
\begin{list}{(\theenumiii)}{\usecounter{enumiii}}  
\item One construct will be a multiplicative group $\bar{K}_{(-1,1)}$
of $2 \times 2$ unimodular real matrix functions $v(\sigma,z_{0})$ of real
variables $\sigma$ and $z_{0}$ such that $-1 < \sigma < 1$, $-1 < z_{0} <
1$ and $\sigma \ne z_{0}$.  These matrices will satisfy certain
differentiability and/or H\"{o}lder conditions and, also, certain
asymptotic conditions at $\sigma=\pm 1,z_{0}$  The specification of
these conditions is still under study.

The subgroup of $\bar{K}_{(-1,1)}$ consisting of all of
its elements which are independent of $z_{0}$ and are analytic functions
of $\sigma$ throughout $-1 < \sigma < 1$ will be denoted by $K_{(-1,1)}$.
Holomorphic continuations of the elements of $K_{(-1,1)}$ into the
$\tau$-plane were used by the authors in previous papers to represent KC
transformations.\cite{17}  The factor group $K_{(-1,1)}/\{I,-I\}$ is a
faithful realization of $K(\Sigma_{\SP})$.

For the current paper, we shall require only that abelian subgroup of
$\bar{K}_{(-1,1)}$ which consists of all of its elements of the form
$\exp{[\sigma_{3} \lambda(\sigma,z_{0})]}$.  The real-valued functions
$\lambda(\sigma,z_{0})$ will be defined in Sec.\ IIG and further
discussed in Secs.\ IIIB, IVA and IVC.

\item The second construct is a homogeneous Hilbert problem (HHP) or
an equivalent linear integral equation whose kernel is simply constructed
from any given $v \in \bar{K}_{(-1,1)}$ and $\bold{F}^{(0)} \in
\Sigma_{\SP}$.  We expect and shall endeavor to prove that (granting its
existence) the solution $\bold{F}$ of the HHP is unique and is also a
member of $\Sigma_{\SP}$, whereupon it will be clear that the HHP defines
a permutation $\Sigma_{\SP} \rightarrow \Sigma_{\SP}$ for each given $v$.
We shall also endeavor to prove that the set of all of these permutations
constitute a group which is isomorphic to $\bar{K}_{(-1,1)}/\{I,-I\}$
and which, of course, we denote by $\bar{K}(\Sigma_{\SP})$.

For the special subsets of $\bar{K}_{(-1,1)}$ and $\Sigma_{\SP}$ which
are to be considered in the current paper, the HHP is expressible as a
simple (non-matrix) Hilbert problem whose general solution is known and
has a finite closed form.  The definition and solution of this simple
Hilbert problem will be given in Sec.\ IIIE.  A definitive version of the
simple Hilbert problem will be given in Sec.\ VIE.
\end{list}

Our formalism is not expected to furnish any new efficacious methods of
generating exact solutions of the Ernst equation.  The advancement of
exact solution productivity is not our aim.

We expect the enlarged group $\bar{K}(\Sigma_{\e})$ to include those
permutations of $\Sigma_{\e}$ which have already been successfully used
to generate exact solutions but which are not in the original group
$K(\Sigma_{\e})$.  For example, we hope to show that our extension
includes the Kramer-Neugebauer involution\cite{18,19,20,21} and those
transformations of C.\ M.\ Cosgrove\cite{22,23,24,25}, D.\ Maison\cite{26}
and G.\ Neugebauer\cite{27} which also involve a non-trivial permutation
of the domain $\{ (r,s) : -1 \le r < s \le 1 \}$.

At an early stage we shall study in detail the precise relation between
our extension of the KC group and the initial value problem for a
\CGPWP.\cite{7,10,28,29}

Our formalism also turns out, as we shall see in this paper, to be
suited for studying the behavior near the {\em axis} (points at which the
exterior product of the Killing vectors becomes zero) of the Ernst
potential $\bold{\E}$.  This will be done in the present paper to provide an
independent method of arriving a a result due to Yurtsever\cite{30} for the
collinear polarization case and we hope to extend our results to the
noncollinear case in another paper of the series.  We contemplate
similar studies on the behavior of the Weyl conform tensor near the
axis, on necessary and sufficient conditions for a horizon to exist on
the axis and (when a horizon exists) continuation of the solution into
the region in which the surface of transitivity of the Killing vectors
is Lorentzian.  We thus intend to reproduce, and possibly augment,
Yurtsever's results\cite{31} on these topics.

So far, we have not provided a clear motivation for enlarging the KC
group.  Why is it necessary to do so?

\subsection{Motivation}
The KC group has been successfully used to generate numerous exact
solutions of the Einstein field equations for stationary axisymmetric
vacuum spacetimes and \CGPWP's.  However, $K(\Sigma_{\e})$ is
``incomplete'' in a sense which will be explained below in (i) and
(ii):
\begin{list}{(\theenumiii)}{\usecounter{enumiii}}  
\item Consider those members of $\Sigma_{\e}$ which have continuous
extensions to the domain
$$\{ (r,s) : -1 \le r < s \le 1 \mbox{ or } -1 < r = s < 1 \}$$
and let $\Sigma_{\e}^{ax}$ be the set of all of these continuous
extensions.  Let $\Sigma_{\e}^{AX}$ be the set of all $\bold{\E}
\in \Sigma_{\e}^{ax}$ such that the restriction of $\bold{\E}$ to
$\{ (r,s) : -1 < r \le s < 1 \}$ is in the differentiability class
$\bold{C}^{2}$.  This is the set for which no spacetime singularities
occur on the {\em axis} $\{ (r,s) : -1 < r = s < 1 \}$ and the
curvature tensor is continuous throughout $\{ (r,s) : -1 < r \le
s < 1 \}$.  The Minkowski space Ernst potential $\bold{\E}^{MS} = -1$ is
in $\Sigma_{\e}^{AX}$ and so are the Ernst potentials for the
Chandrasekhar-Xanthopoulos\cite{32} and Hoenselaers-Ernst\cite{33}
\CGPWP\ interaction
regions (which are respectively isometric to different subregions
of the Kerr metric ergosphere).  On the other hand, the Ernst
potentials for the Nutku-Halil\cite{34} spacetime and for any other \CGPWP\
spacetime which has curvature singularities on the axis are in
$\Sigma_{\e}-\Sigma_{\e}^{AX}$.

Now let $\bold{\E}^{(0)} \in \Sigma_{\e}^{AX}$ and let $\bold{\E}^{(0)}
\rightarrow \bold{\E}$ under the KC transformation corresponding
to any given member $v$ of the analytic matrix group $K_{(-1,1)}$.
Then methods which the authors have previously used for stationary
axisymmetric vacuum spacetimes can be applied here to prove that
$\bold{\E}$ is also in $\Sigma_{\e}^{AX}$ and that the following {\em
axis relation} holds:
\begin{equation}
\bold{\E}(z) = i [v^{1}_{\; 1}(z) \bold{\E}^{(0)}(z) +
i v^{1}_{\; 2}(z)] [v^{2}_{\; 1}(z) \bold{\E}^{(0)}(z) + i
v^{2}_{\; 2}(z)]^{-1}
\end{equation}
where $\bold{\E}(z) := \bold{\E}(z,z)$ and $-1 < z < 1$.\cite{35,36}

It follows that no member of $\Sigma_{\e}-\Sigma_{\e}^{AX}$ can
be obtained from a member of $\Sigma_{\e}^{AX}$ (and vice versa)
by applying an element of $K(\Sigma_{\e})$.

\item Let $\Sigma_{\e}^{an}$ be the set of all $\bold{\E} \in \Sigma_{\e}$
such that the initial data $\bold{\E}_{3}(r) = \bold{\E}(r,1)$ and
$\bold{\E}_{2}(s) = \bold{\E}(-1,s)$ are analytic functions throughout
$-1 < r < 1$ and $-1 < s < 1$, respectively.  The set $\Sigma_{\e}^{an}$
is clearly not empty since the Ernst potentials for the Kasner metrics and
for every published exact \CGPWP\ solution with noncollinear polarizations
lie in this set.  The set $\Sigma_{\e}-\Sigma_{\e}^{an}$ is also not
empty even if we restrict ourselves to Ernst potentials which are
$C^{\infty}$ throughout their domains.  The basic reason for this is
that the Ernst equation for $\Sigma_{\e}$ is hyperbolic (unlike the
Ernst equation for the stationary axisymmetric case).  Now, a
procedure like one previously used by the authors for stationary
axisymmetric spacetimes can be used to prove that each element of
$K(\Sigma_{\e})$ maps $\Sigma_{\e}^{an}$ onto $\Sigma_{\e}^{an}$.\cite{35,36}
Therefore, each element of $K(\Sigma_{\e})$ maps $\Sigma_{\e}-
\Sigma_{\e}^{an}$ onto $\Sigma_{\e}-\Sigma_{\e}^{an}$.
\end{list}

In summary, $K(\Sigma_{\e})$ is not transitive.  Therefore,
$K(\Sigma_{\SP})$ is not transitive.  This is our principal motivation
for extending the KC group.  We expect $\bar{K}(\Sigma_{\SP})$ to
be transitive and we shall try to prove this in a future paper.
In other words, we hope to prove that our extension can be used
in principle to generate any $\bold{\E} \in \Sigma_{\e}$ from any known
given $\bold{\E}^{(0)} \in \Sigma_{\e}$ (e.g., the Minkowski space Ernst
potential $\bold{\E}^{MS} = -1$).  The proof will have at least two
difficult parts:
\begin{enumerate}
\item A proof that the solution of the HHP, granted its existence,
yields a member $\bold{F}$ of $\Sigma_{\SP}$.
\item A proof that the solution of the HHP always exists and has
the requisite differentiability properties.
\end{enumerate}

Note that the conjecture that $\bar{K}(\Sigma_{\SP})$ is transitive
resembles a conjecture due to Geroch\cite{4} for the stationary axisymmetric
vacuum spacetimes.  An explication of the Geroch conjecture was
formulated and proven by the authors.\cite{36}  An essential part of the
proof was establishing the existence of a solution of the HHP used
to effect KC transformations, i.e., a proof that the KC transformations
actually exist.  It is easy to forget that one must prove existence
of the group.

There are two more objectives which are appropriate to mention at
this point.  One of these concerns the set $\Sigma_{\e}^{ax}$ which
we defined above.  We conjecture and hope to prove in a later paper
that the element of $\bar{K}(\Sigma_{\e})$ corresponding to any
given $v \in \bar{K}_{(-1,1)}$ maps $\Sigma_{\e}^{ax}$ onto
$\Sigma_{\e}^{ax}$ if and only if $v(\sigma,z_{0})$ is independent
of $z_{0}$.  This is proven in Sec.\ IVC for those subsets of
$\Sigma_{\e}$ and $\bar{K}_{(-1,1)}$ which are treated in the
current paper.

Another objective concerns the set $\Sigma_{\e}^{\CW}$ of all
members of $\Sigma_{\e}$ for which the colliding wave conditions
hold.  More precisely, it concerns certain substantial subsets
of $\Sigma_{\e}^{\CW}$ which are defined and discussed in Sec.\ VE
of this paper.  We have proven (unpublished as yet) that each of
these subsets of $\Sigma_{\e}^{\CW}$ is mapped onto itself by the
KC transformation corresponding to any given $v \in K_{(-1,1)}$
which is holomorphic on the closed interval $-1 \le \sigma \le 1$.
In a later paper, we hope to extend this result to
$\bar{K}(\Sigma_{\SP})$ elements corresponding to members $v$ of
$\bar{K}_{(-1,1)}$ which are not necessarily analytic but which
do have certain ``well behaved'' extensions to the union of the
intervals $-1 \le \sigma < z_{0}$ and $z_{0} < \sigma \le 1$.
For the special case which is treated in the current paper, this
objective is fulfilled in Sec.\ VF.

\subsection{The Principal Results for the Weyl Metrics}
Before we launch into formal developments, we shall review the
definition of $\E$ and then give a preview of the main results
of this paper.  Let the line
element $(+++-)$ be expressed as
\begin{equation}
ds^{2} = dx^{a} dx^{b} g_{ab}(u,v) + 2 du dv g_{34}(u,v) ,
\end{equation}
where $a=1,2$ and $b=1,2$, where $x^{1},x^{2}$ are the ignorable
coordinates and where $x^{3} = u$ and $x^{4} = v$.  Introduce a
two dimensional duality operator $*$ on $1$-forms which are linear
combinations of $du$ and $dv$:
\begin{equation}
* du = du \; , \; * dv = - dv \; .
\end{equation}
Thus $**=1$.  The Ernst potential $\E$ is defined by
\begin{equation}
\E = f + i \chi \; , \; f := - g_{22} \; ,
\end{equation}
where the {\em twist potential} $\chi$ is defined as the integral of
\begin{equation}
\rho d\chi = * f^{2} d\omega ,
\end{equation}
where
\begin{equation}
\omega := g_{12}/g_{22} \; , \; \rho := \sqrt{\det{g_{ab}}} \; .
\end{equation}

{}From the above definition, $\E$ satisfies the Ernst equation
\begin{equation}
f \; d(*\rho d\E) = \rho d\E (*d\E) .
\end{equation}
Conversely, if a solution of Eq.\ (1.9) is given so that its
real part $f$ is negative in value, then the metric components
$g_{ab}(u,v)$ and $g_{34}(u,v)$ can be computed from $\E$ by
employing Eqs.\ (1.6), (1.7) and (1.8) and other equations which
will be given in Sec.\ VB.  Unless we explicitly state otherwise,
we shall fix the various constants of integration by adopting
conventions so that $\E, g_{ab}$ and $g_{34}$ have the values
$-1, \delta_{ab}$ and $-1$, respectively, at $u=v=0$.

For the Weyl vacuum metrics, the metric
components $g_{ab}$ in the line element (1.4) are expressible in the
conventional forms
\begin{equation}
g_{12} = 0 \; , \; g_{22} = e^{2\psi} \; , \; g_{11} = \rho^{2}
e^{-2\psi} \; ,
\end{equation}
whereupon
\begin{equation}
\E = - \exp{(2\psi)} .
\end{equation}
Note that the above usage of `$\psi$' differs from that in some previous
papers by the authors and others.\cite{10}

We recall that the vacuum field equations imply that\cite{37}
\begin{equation}
\rho(u,v) = \frac{1}{2} [s(v)-r(u)] ,
\end{equation}
where
\begin{equation}
\dot{r}(u) > 0 \mbox{ for } 0 < u < u_{0} \; , \; \dot{s}(v) < 0
\mbox{ for } 0 < v < v_{0} \; ,
\end{equation}
and
\begin{equation}
\dot{r}(0) = \dot{s}(0) = 0 .
\end{equation}
By appropriate scalings and choice of the arbitrary constant in $r(u)$
and $s(v)$, we impose the conventions:
\begin{equation}
r(0)=-1 \; , \; s(0)=1 \; , \; \psi(0,0) = 0 \; , \; r(u_{0})=1 \; , \;
s(v_{0})=-1 \; .
\end{equation}
The domain of the chart in the region of interaction is
\begin{equation}
{\rm IV} := \{ (u,v):0 \le u < u_{0} \; , \; 0 \le v < v_{0} \; , \;
-1 \le r(u) < s(v) \le 1
\} .
\end{equation}

In a previous paper on the initial value problem for \CGPWP's,
I.\ Hauser and F.\ J.\ Ernst\cite{7} introduced the notations
\begin{equation}
\psi_{3}(u) := \psi(u,0) \; , \; \psi_{2}(v) := \psi(0,v)
\end{equation}
for the initial values of $\psi$, i.e., for the values in the plane wave
regions III and II.  An important role in our form of the solution of
the initial value problem was played by the Abel transforms
[$\dot{\bold{\psi}}_{3}(r) := d\bold{\psi}(r)/dr$]\cite{7,10}
\begin{equation}
g_{3}(\sigma) := \int_{-1}^{\sigma} dr \frac{\dot{\bold{\psi}}_{3}(r)}
{\sqrt{\sigma-r}} \; , \;
g_{2}(\sigma) := \int^{1}_{\sigma} ds \frac{\dot{\bold{\psi}}_{2}(s)}
{\sqrt{s-\sigma}} \; ,
\end{equation}
where we are following a practice, which was discussed in detail in a
previous paper\cite{28}, of employing corresponding boldface letters for
functions with the domain
\begin{equation}
\DIV := \{ (r,s): -1 \le r < s \le 1 \}
\end{equation}
such that, for example,
\begin{equation}
\bold{\psi}_{3}(r(u)) := \psi_{3}(u) \; , \;
\bold{\psi}_{2}(s(v)) := \psi_{2}(v) \; , \;
\bold{\psi}(r(u),s(v)) := \psi(u,v) \; .
\end{equation}

Now let us introduce ($-1 < \sigma < 1$)
\begin{equation}
\G_{3}(\sigma) := \sqrt{1-\sigma} g_{3}(\sigma) \; , \;
\G_{2}(\sigma) := \sqrt{1+\sigma} g_{2}(\sigma) \; ,
\end{equation}
and, for any given real number $z_{0}$ such that $-1 < z_{0} < 1$, let
\begin{equation}
\G(\sigma,z_{0}) := \left\{ \begin{array}{ccc}
\G_{3}(\sigma) & \mbox{if} & -1 < \sigma < z_{0} , \\
\G_{2}(\sigma) & \mbox{if} & z_{0} < \sigma < 1 ,
\end{array} \right.
\end{equation}
and
\begin{equation}
\lambda(\sigma,z_{0}) = - \frac{1}{\pi} \int_{-1}^{1} d\sigma'
\frac{\G(\sigma',z_{0})}{\sigma' - \sigma}
\mbox{ (principal value)}
\end{equation}
for $-1 < \sigma < 1$ and $\sigma \ne z_{0}$.  Then the general
solution for $\bold{\psi}$ on the domain $\DIV$ is expressible
in the form
\begin{equation}
\bold{\psi}(r,s) = \frac{1}{\pi} \int_{r}^{s} d\sigma
\frac{\lambda(\sigma,z_{0})}{\sqrt{(\sigma-r)(s-\sigma)}}
= \frac{1}{\pi} \int_{-\frac{\pi}{2}}^{\frac{\pi}{2}} d\theta
\lambda(z+\rho \sin \theta,z_{0}) \; ,
\end{equation}
where $r < z_{0} < s \; , \; z = \frac{1}{2}(s+r) \; , \; \rho =
\frac{1}{2}(s-r)$.
The above expression for $\bold{\psi}(r,s)$ is the principal result
of this paper.  Note that the integral is independent of $z_{0}$.

As regards the connection with our new group realization, consider the
additive group of all of the functions $\lambda(\sigma,z_{0})$
corresponding to all possible initial data functions $\bold{\psi}_{3}$
and $\bold{\psi}_{2}$.  Then the corresponding multiplicative group of
all $2 \times 2$ matrix functions
\begin{equation}
v(\sigma,z_{0}) = \exp{[\sigma_{3} \lambda(\sigma,z_{0})]}
\end{equation}
is a subgroup of the group realization.  Here, $\sigma_{3}$ is the
usual Pauli spin matrix.

Further study of the principal result (1.24) shows that the continuous
extension to the axis points $(r,s)=(z,z), -1 < z < 1$, exists if and
only if
\begin{equation}
\G_{3}(\sigma) = \G_{2}(\sigma) \mbox{ (necessary condition for a
horizon to exist)},
\end{equation}
whereupon $\lambda(\sigma,z_{0})$ is independent of $z_{0}$ and
\begin{equation}
\lambda(\sigma) := \lambda(\sigma,z_{0}) = \bold{\psi}(\sigma,\sigma) .
\end{equation}
Thus the connection between the axis values and the initial values of
the potential is established.  When the axis is not accessible, one
obtains the result that
\begin{equation}
\bold{\psi}(r,s) + \frac{1}{\pi} [\G_{3}(z)-\G_{2}(z)] \ln{\rho}
\end{equation}
has a continuous extension to the axis points $-1 < z < 1$.  This is
consistent with Yurtsever's result.\cite{30}

%
\setcounter{equation}{0}

\section{Spectral Potentials $\xi$ and Functions $\lambda$}

\subsection{Preliminaries}

We shall be concerned in this paper only with the real Ernst potentials
$\E = -\exp(2\psi)$ or, equivalently,
with the potentials $\psi$.  Corresponding to each $\psi$, there is a
gauge of potentials $\xi$ which depend on a complex (spectral)
parameter $\tau$ as well as the coordinates $u$ and $v$, which have
the property that $\xi(u,v,\tau)$ is holomorphic in a neighborhood of
$\tau=\infty$ and which satisfy $\xi(u,v,\infty)=\psi(u,v)$.  Spectral
potentials which are closely related to $\xi$ were used in the
treatment of the initial value problem for $\psi$ by I.\ Hauser and
F.\ J.\ Ernst\cite{7,10}, and the spectral potentials $\xi$ will play a
similar role in our exposition of the KC group extension for Weyl metrics.

The present section will be devoted to the concept of the
$\xi$-potential and to some key properties of $\xi$.  One important
result will be a new form of the general solution for $\xi$
corresponding to any given initial data and a byproduct of this result
will be the expression (1.24) for $\bold{\psi}$.

\begin{description}
\item[Premise:]  In this section and in the next two,
we shall assume only that the
initial data functions $r(u), \psi_{3}(u)$ and $s(v), \psi_{2}(v)$ are
in the differentiability class ${\bf C}^{1}$ over the intervals $0 \le
u < u_{0}$ and $0 \le v < v_{0}$, respectively, and that the inequality
conditions (1.13) hold.  The following known theorem then holds.
\item[Theorem:]  There exists exactly one function $\psi$ whose domain
is IV such that the partial derivatives $\psi_{u}(u,v),
\psi_{v}(u,v)$ and $\psi_{uv}(u,v)$ exist and are continuous
throughout IV and such that
$$\psi(u,0) = \psi_{3}(u) \; , \; \psi(0,v) = \psi_{2}(v) \; ,$$
and $d * (\rho d\psi) = 0$.
\end{description}

\subsection{The Sets ${\bf D}_{IV}, \bar{D}_{IV}, \sigma^{\pm},
(r,s)^{\pm}$, and dom $\mu$}

We shall follow the practice of a previous paper\cite{7} and regard the
domain $\DIV$ which was defined by Eq.\ (1.19) as a two-dimensional
manifold for which the mapping of $\DIV$ onto IV given by
$$\left(r(u),s(v)\right) \rightarrow (u,v)$$
is a chart.  Therefore, concerning the function $\bold{\psi}$ which
was defined by Eq.\ (1.20) and has $\DIV$ as its domain,
$d\bold{\psi}$, $d*d\bold{\psi}$ and $d^{2}\bold{\psi}$ are well
defined (in terms of the chart) even though the partial derivatives
$\bold{\psi}_{r}(r,s)$ and $\bold{\psi}_{s}(r,s)$ do not generally
exist at $r=-1$ and at $s=1$, respectively.  The following extensions
of $\DIV$ will be used later.

\begin{description}
\item[Definitions:]  Let
\begin{subequations}
\begin{equation}
{\bf D}_{IV} := \DIV \cup \{ (z,z):-1<z<1 \}
\end{equation}
and
\begin{equation}
\bar{D}_{IV} := {\bf D}_{IV} \cup \{(-1,-1),(1,1)\} .
\end{equation}
\end{subequations}
\end{description}

The Ernst equation for the Weyl metrics is equivalent to the linear
equation
\begin{equation}
d(\rho*d\bold{\psi}) = 0 ,
\end{equation}
where $*dr=dr$ and $*ds=-ds$.  A solution of Eq.\ (2.2) which is
expressible as a function of $r$ times a function of $s$ is
$\mu(r,s,\tau)^{-1}$ where
\begin{equation}
\mu(r,s,\tau) = \left[(\tau-r)(\tau-s)\right]^{\frac{1}{2}}
\end{equation}
and $\tau$ is a separation parameter which will be assigned complex
values.  We shall not use the entire Riemann surface (for fixed
$(r,s)$ in \DIV) of the analytic function of $\tau$ on the right side
of Eq.\ (2.3).  The domain of $\mu$ and various related concepts are
defined below.

\begin{description}
\item[Definitions:] \
\begin{list}{(\theenumiii)}{\usecounter{enumiii}}  
\item We let $C$ denote the extended complex plane, $[r,s]$ denote the
closed real axis interval whose end points are $r$ and $s (r \le s)$
and $]r,s[$ denote the open real axis interval whose end points are
$r$ and $s (r < s)$.
\item Corresponding to each $\sigma$ in $]-1,1[$, let $\sigma^{+}$ and
$\sigma^{-}$ denote the sets of all sequences
$$\sigma+w_{n} \mbox{ and } \sigma-w_{n} \; (n=1,2,\ldots) \; ,$$
respectively, such that $\Im{w_{n}} > 0$ and $w_{n} \rightarrow 0$ as
$n \rightarrow \infty$.  For any given $(r,s) \in \DIV$, let the sets
$(r,s)^{+}$ and $(r,s)^{-}$ be
$$(r,s)^{\pm} := \{ \sigma^{\pm}:r < \sigma < s \} \; .$$
\item We shall employ that branch of the Riemann surface for (2.3) such
that
\begin{eqnarray}
\lefteqn{ \dom{\mu} := \{ (r,s,t):(r,s) \in \bar{D}_{IV} } \nonumber \\
& & \mbox{and }
	\tau \in \left[ \left(C-]r,s[\right)
	\cup (r,s)^{+} \cup (r,s)^{-} \right] \} ,
\end{eqnarray}
such that, for fixed $(r,s)$, $\mu(r,s,\tau)$ is a holomorphic function
of $\tau$ throughout $C-[r,s]$ and satisfies
\begin{equation}
\mu(r,s,\tau)/\tau = 1 \mbox{ when } \tau = \infty \; ,
\end{equation}
such that, for fixed $(r,s)$ in $\DIV$ and $r < \sigma < s$,
\begin{eqnarray}
\mu(r,s,\sigma^{\pm}) & := & \lim_{w \rightarrow 0} \mu(r,s,\sigma \pm w)
\nonumber \\
& = & \pm i \sqrt{(\sigma-r)(s-\sigma)} \; \;  (\Im{w} > 0) \; ,
\end{eqnarray}
and such that
$$\mu(r,s,r) = \mu(r,s,s) = 0.$$
\item For fixed $(r,s)$ in $\DIV$, there is an obvious one-to-one
mapping $\tau \rightarrow p(\tau)$ of the domain of $\mu(r,s,\tau)$
onto the closure of one of the sheets of the Riemann surface for
$[(\tau-r)(\tau-s)]^{1/2}$ such that the points $p(\sigma^{\pm})$
constitute the two open line segments which form the bounding edges of
the sheet and have common end points $p(r)$ and $p(s)$.  We adopt that
topology for the domain of
$\mu(r,s,\tau)$ such that the mapping $\tau \rightarrow
p(\tau)$ is a homeomorphism.  The disjoint lines $(r,s)^{+}$ and
$(r,s)^{-}$ will be called the {\em upper} and {\em lower lips},
respectively, of the real axis interval $]r,s[$ in the $\tau$-plane.
\item For any point $\tau$ in the union of $C,(-1,1)^{+}$ and
$(-1,1)^{-}$, let $\mu(\tau)$ denote that function whose domain is
$$\dom{\mu(\tau)} := \{ (r,s): -1 \le r \le s \le 1 \mbox{ and }
\mu(r,s,\tau) \ne 0 \}$$
and whose values are $\mu(\tau)(r,s) := \mu(r,s,\tau)$.  The function
represented by the ratio $\tau/\mu(\tau)$ is understood to have the
value $1$ when $\tau=\infty$.
\end{list}
\item[Properties of $\mu$:]
It is clear that $\mu$ is continuous throughout its domain and that
\begin{equation}
\mu(z,z,\tau) = \tau - z
\end{equation}
for all $z$ in $[-1,1]$.  Also, $\tau/\mu(\tau)$ is an analytic
function of $(r,s)$ throughout its domain.  When $\tau=\sigma$ is real
and $-1 < \sigma < 1$, the domain of $\mu(\sigma^{+})$ is the same as
the domain of $\mu(\sigma^{-})$, but the domain of $\mu(\sigma)$
consists of two disjoint subregions of $\bar{D}_{IV}$.  These domains
are illustrated in Fig.\ 1.
\end{description}

\subsection{The Plemelj Theorem\cite{38}}

Domains like that of $\mu$ will frequently be used for functions which
are defined by integrals of Cauchy's type.  For example, let
$f(\sigma)$ be any real valued function defined for all $\sigma$ on an
open interval $a < \sigma < b$ and summable on $[a,b]$.  Suppose $f$
also obeys a H\"{o}lder condition on every closed subinterval $[c,d]$
such that $a < c < d < b$; i.e., there exist positive real numbers
$M(c,d)$ and $\nu(c,d)$ such that $0 < \nu(c,d) \le 1$ and
$$\left| f(\sigma_{2})-f(\sigma_{1}) \right| \le
M(c,d) \left| \sigma_{2}-\sigma_{1} \right|^{\nu(c,d)}$$
for all $\sigma_{1},\sigma_{2}$ in $[c,d]$.  The number $\nu(c,d)$ is
called the {\em index} of the H\"{o}lder condition obeyed by $f$ on
$[c,d]$.

Now let $F$ by that function whose domain is
$$\left( C - [a,b] \right) \cup (a,b)^{+} \cup (a,b)^{-}$$
and which is defined by the integral
\begin{equation}
F(\tau) := \frac{1}{\pi} \int_{a}^{b} d\sigma
\frac{f(\sigma)}{\sigma - \tau} ,
\end{equation}
where, for all $a < \sigma < b$,
\begin{equation}
F(\sigma^{\pm}) := \lim_{w \rightarrow 0} F(\sigma \pm w) \; \;
(\Im{w} > 0) \; .
\end{equation}
A slight generalization of a theorem due to I.\ Plemelj asserts the
following:

\begin{description}
\item[The Plemelj Theorem:] \

\begin{list}{(\theenumiii)}{\usecounter{enumiii}}
\item
The function $F$ exists and so does
\begin{equation}
\bar{F}(\sigma) := \frac{1}{\pi} \int_{a}^{b} d\sigma'
\frac{f(\sigma')}{\sigma' - \sigma} \; \;
(a < \sigma < b) \; .
\end{equation}
The principal value is always to be understood in integrals like the
above.  The function $F$ is holomorphic throughout $C-[a,b]$,
satisfies $F(\infty)=0$, and is continuous throughout the union
of $C-[a,b]$, $(a,b)^{+}$ and $(a,b)^{-}$.
\item
The function $\bar{F}$ obeys a H\"{o}lder condition on any given
closed subinterval $[c,d]$ of $]a,b[$.  If $\nu(c,d)$ is the index
for $f$, then the index for $\bar{F}$ is also $\nu(c,d)$ if $\nu(c,d) < 1$
and is $1-\epsilon$ (for an arbitrarily small $\epsilon > 0$) if
$\nu(c,d)=1$.
\item
For all $a < \sigma < b$,
\begin{equation}
F(\sigma^{\pm}) = \bar{F}(\sigma) \pm i f(\sigma) \; .
\end{equation}
\end{list}
\end{description}

In particular, we shall be employing Cauchy type integrals involving
the functions $\G_{3}(\sigma)$ and $\G_{2}(\sigma)$ which are
defined by Eqs.\ (1.18) and (1.21).  The following pertinent theorems
have been proven by I.\ Hauser and F.\ J.\ Ernst.\cite{7,10}

\begin{description}
\item[Theorems:] \
\begin{enumerate}
\item The functions $g_{3}(\sigma)/\sqrt{c-\sigma}$ and
$g_{2}(\sigma)/\sqrt{\sigma-c}$ are summable on $[-1,c]$ and $[c,1]$,
respectively, where $c$ is any real number such that $-1 < c < 1$.
\item Also, $g_{3}(\sigma)$ and $g_{2}(\sigma)$ both obey H\"{o}lder
conditions of index $\frac{1}{2}$ on any given closed subinterval of
the open interval $-1 < \sigma < 1$.
\end{enumerate}
\end{description}

The above two statements also hold if $g_{3}$ and $g_{2}$ are replaced
by $\G_{3}$ and $\G_{2}$, respectively.

\subsection{The Special Gauge of Spectral Potentials $\xi$}

One readily proves that
\begin{equation}
\rho * d[1/\mu(\tau)] = - d[(\tau-z)/\mu(\tau)] .
\end{equation}
Also, $\omega * (\chi) = - (*\omega)\chi$ for any $1$-forms
$\omega$ and $\chi$ in $\DIV$.
Therefore, from Eq.\ (2.2)
\begin{equation}
d \left[ \frac{(\tau-z-\rho*) d\bold{\psi}}{\mu(\tau)} \right]
= 0 .
\end{equation}
The domain (for given $\tau$) of the expression in brackets in the
above equation is the intersection of the domain of $\tau/\mu(\tau)$
wtih $\DIV$.

For now, let us avoid the complications which arise when $\tau$ is real
and lies on $[-1,1]$ (Fig.\ 1 illustrates some of these complications)
by {\em temporarily} confining our discussion to the domain
\begin{equation}
\DIV \times \left( C - [-1,1] \right) =
\left\{ (r,s,\tau):(r,s) \in \DIV \mbox{ and } \tau \in
C - [-1,1] \right\} .
\end{equation}
Equation (2.13) clearly implies the existence of infinitely many
functions $\bold{\xi}$ which have the above domain such that, if we let
$\bold{\xi}(\tau)$ denote that function whose domain is $\DIV$ and
whose values are
$$\bold{\xi}(\tau)(r,s) := \bold{\xi}(r,s,\tau),$$
then $\bold{\xi}(\tau)$ is a holomorphic function of $\tau$ throughout
$C - [-1,1]$, and $d\bold{\xi}(\tau)$ exists and satisfies
\begin{equation}
d\bold{\xi}(\tau) = \left[ \frac{\tau-z-\rho *}{\mu(\tau)} \right]
d\bold{\psi}
\end{equation}
and
\begin{equation}
\bold{\xi}(\infty) = \bold{\psi} .
\end{equation}
There is exactly one solution $\bold{\xi}$ of the above conditions and
equations for each choice of
\begin{equation}
\bold{\xi}_{1}(\tau) := \bold{\xi}(-1,1,\tau)
\end{equation}
such that $\bold{\xi}_{1}(\tau)$ is holomorphic throughout $C - [-1,1]$
and satisfies $\bold{\xi}_{1}(\infty) = 0$.  Note that
$\bold{\xi}_{1}(\tau)$ are the values of $\bold{\xi}(\tau)$ at the
collision points of the colliding plane waves.

Now, what we require for the purpose of this paper is a very special
choice of $\bold{\xi}_{1}(\tau)$, viz., one which maximizes the domain
of holomorphy of $\bold{\xi}(r,s,\tau)$ in the $\tau$-plane.  This
selection principle and our previous experiences with specific
examples led us to seek a choice of $\bold{\xi}_{1}(\tau)$ such that,
for fixed $(r,s)$, $\bold{\xi}(r,s,\tau)$ has a holomorphic extension
in the $\tau$-plane to the domain $C-[r,s]$.  This is, of course, the
same domain of holomorphy as that of $\mu(r,s,\tau)$ and is the
maximal domain of holomorphy one can generally hope to achieve.

It turned out that our selection principle for $\bold{\xi}_{1}(\tau)$
resulted in a more interesting structure than the one originally
envisioned.  We were inevitably led to a one-parameter family of
choices of $\bold{\xi}_{1}(\tau)$:
$$\bold{\xi}_{1}(\tau,z_{0}) \mbox{ where } -1 < z_{0} < 1$$
such that the corresponding $\bold{\xi}$-potential
$\bold{\xi}(r,s,\tau,z_{0})$ has a holomorphic extension to $C-[r,s]$
{\em provided that} one restricts $(r,s,z_{0})$ to the range $-1 \le r
< z_{0} < s \le 1$.

For example, consider the Kasner metrics for which $\bold{\psi}$ is
given by
\begin{equation}
\bold{\psi}^{K}(r,s) = \left( \frac{1-n}{2} \right) \ln \rho
\end{equation}
where $n$ is any real number and the Weyl canonical coordinates $z$
and $\rho$ are
\begin{equation}
z = \frac{1}{2}(s+r) \; , \; \rho = \frac{1}{2}(s-r) \; .
\end{equation}
We found the corresponding one-parameter family of spectral potentials
\begin{equation}
\bold{\xi}^{K}(r,s,\tau,z_{0}) = \left( \frac{1-n}{2} \right)
\ln \left[ \frac{2(\tau-z_{0}) \rho}{\tau-z+\mu(r,s,\tau)} \right]
\end{equation}
where the parameter $z_{0}$ has the range
$$-1 < z_{0} < 1$$
and where the cut is the union of $[r,z_{0}]$ and $[z_{0},s]$, and the
value at $\tau = \infty$ is $\bold{\psi}^{K}(r,s)$.  An alternative
form is
\begin{equation}
\bold{\xi}^{K}(r,s,\tau,z_{0}) = \bold{\psi}^{K}(r,s)
+ (n-1) \ln \left\{ \frac{1}{2} \left[ \left(
\frac{\tau-r}{\tau-z_{0}} \right)^{\frac{1}{2}} + \left(
\frac{\tau-s}{\tau-z_{0}} \right)^{\frac{1}{2}} \right] \right\}
\end{equation}
with branch cuts $[r,z_{0}]$ and $[z_{0},s]$ for the respective square
roots both of which have unit value at $\tau=\infty$.  For fixed
$(r,s,z_{0})$, it is clear that the above
$\bold{\xi}^{K}(r,s,\tau,z_{0})$ is a holomorphic function of $\tau$
throughout $C-[r,s]$ {\em provided that} $(r,s,z_{0})$ is restricted
to the range
$$-1 \le r < z_{0} < s \le 1.$$
This restriction presents no problem since one can cover any given
point $(r,s)$ in \DIV simply by selecting $z_{0}$ so that $r < z_{0} <
s$.

There are other pertinent properties of the above one-parameter family
of spectral potentials.  Observe that $\bold{\xi}^{K}(r,s,\tau,z_{0})$
is a continuous function of $\tau$ over the domain
$$(C-]r,s[) \cup (r,z_{0})^{+} \cup (r,z_{0})^{-} \cup (z_{0},s)^{+}
\cup (z_{0},s)^{-} \; ,$$
and that the singularity at $\tau=z_{0}$ is logarithmic.

We have found in fact that there exists a similar one-parameter gauge
of spectral potentials for any given $\bold{\psi}$.  This gauge will
play a vital role in our extension of the KC group and will be called
the {\em special gauge} of spectral potentials $\bold{\xi}$.  The
first step in formulating the definition of the special gauge will be
to introduce some domains which are used in its definition.

\subsection{The Sets ${\bf D}(r,s,z_{0}), {\bf D}(\tau,z_{0}),
{\bf D}$, and the Special Gauge of Spectral Potentials $\xi$}

\begin{description}
\item[Definitions:] \
\begin{list}{(\theenumiii)}{\usecounter{enumiii}}

\item
For each choice of $(r,s,z_{0})$ such that
$$-1 \le r < z_{0} < s \le 1,$$
let
\begin{eqnarray}
\lefteqn{\bold{D}(r,s,z_{0}) :=
\{ \tau: \tau \mbox{ is any point in the union of }} \nonumber \\
& & C-[r,s], (r,z_{0})^{+}, (r,z_{0})^{-}, (z_{0},s)^{+}, (z_{0},s)^{-},
\nonumber \\
& & \mbox{ the singlet set $\{ r \}$ if $r \ne -1$}, \nonumber \\
& & \mbox{ and the singlet set $\{ s \}$ if $s \ne 1$} \} .
\end{eqnarray}

\item
For each choice of $(\tau,z_{0})$ such that $-1 < z_{0} < 1$ and
\begin{equation}
\tau \in \left[ (C-\{ z_{0} \}) \cup (-1,z_{0})^{+} \cup
(-1,z_{0})^{-} \cup (z_{0},1)^{+} \cup (z_{0},1)^{-} \right] ,
\end{equation}
let $\bold{D}(\tau,z_{0})$ be defined as follows for various ranges of
$\tau$:
\begin{subequations}
\begin{eqnarray}
\bold{D}(\tau,z_{0}) & := & \{(r,s):-1 \le r < z_{0} < s \le 1 \} \nonumber \\
& & \mbox{ if } \tau \in (C-[-1,1]) , \\
\bold{D}(-1,z_{0}) & := & \{(r,s):-1 < r < z_{0} < s \le 1 \} , \\
\bold{D}(1,z_{0}) & := & \{(r,s):-1 \le r < z_{0} < s < 1 \} , \\
\bold{D}(\sigma^{\pm},z_{0}) & := & \{(r,s):-1 \le r < z_{0} < \sigma
\le s \le 1 \} \nonumber
\\ & & \mbox{ if } z_{0} < \sigma < 1 , \\
\bold{D}(\sigma,z_{0}) & := & \{(r,s):-1 \le r < z_{0} < s \le \sigma
\} \nonumber \\
& & \mbox{ if } z_{0} < \sigma < 1 , \\
\bold{D}(\sigma^{\pm},z_{0}) & := & \{(r,s):-1 \le r \le \sigma
< z_{0} < s \le 1 \} \nonumber \\
& & \mbox{ if } -1 < \sigma < z_{0} , \\
\bold{D}(\sigma,z_{0}) & := & \{(r,s):\sigma \le r < z_{0} < s \le 1
\} \nonumber \\
& & \mbox{ if } -1 < \sigma < z_{0} .
\end{eqnarray}
\end{subequations}
Figure 2 depicts $\bold{D}(\tau,z_{0})$ for the cases (2.24a), (2.24b)
and (2.24c).  Figure 3 depicts the cases (2.24d) and (2.24e).  Figure
4 depicts the cases (2.24f) and (2.24g).  It is instructive to compare
Figs.\ 3 and 4 with Fig.\ 1.

\item
Let
\begin{equation}
\bold{D} := \{ (r,s,\tau,z_{0}): -1 \le r < z_{0} < s \le 1
\mbox{ and } \tau \in \bold{D}(r,s,z_{)}) \} .
\end{equation}
Equivalently,
\begin{eqnarray}
\bold{D} & := & \{ (r,s,\tau,z_{0}): -1 < z_{0} < 1, \tau
\mbox{ is any point} \\
& & \mbox{in the set (2.23) and } (r,s) \in
\bold{D}(\tau,z_{0}) \} . \nonumber
\end{eqnarray}

\item
If $\bold{\xi}$ denotes a function whose domain is $\bold{D}$, then
corresponding to each choice of $(\tau,z_{0})$ such that $-1 < z_{0} <
1$ and $\tau$ is in the set (2.23), $\bold{\xi}(\tau,z_{0})$ will
denote that function which has domain $\bold{D}(\tau,z_{0})$ and
values
\begin{equation}
\bold{\xi}(\tau,z_{0})(r,s) := \bold{\xi}(r,s,\tau,z_{0}) .
\end{equation}
\end{list}
\end{description}

We next define the special gauge of spectral potentials $\bold{\xi}$.
The definition is given without regard to the questions of existence
and uniqueness.  The existence and uniqueness of $\bold{\xi}$ for
given $\bold{\psi}$ will be covered in Sec.\ III.  In the meantime,
existence will be granted.
\begin{description}
\item[Definition:]
For given $\bold{\psi}$, we shall henceforth let $\bold{\xi}$ denote a
function whose domain is $\bold{D}$ and which satisfies the conditions
(i) to (iv) given below.
\begin{list}{(\theenumiii)}{\usecounter{enumiii}}
\item
The function $\bold{\xi}(\tau,z_{0})$ obeys the following integral
equation for all $(r,s)$ and $(a,b)$ in $\bold{D}(\tau,z_{0})$:
\begin{equation}
\bold{\xi}(r,s,\tau,z_{0}) = \bold{\xi}(a,b,\tau,z_{0})
+ \int^{(r,s)}_{(a,b)} \left(
\frac{\tau - z' - \rho' \; *}{\tau - z' + \rho' \; *}
\right)^{\frac{1}{2}} d\bold{\psi}(r',s')
\end{equation}
where note that
\begin{equation}
\left(
\frac{\tau - z' - \rho' \; *}{\tau - z' + \rho' \; *}
\right)^{\frac{1}{2}} = \frac{\tau - z' - \rho' \; *}{\mu(r',s',\tau)} ,
\end{equation}
since $(\tau - z' - \rho' \; *)(\tau - z' + \rho' \; *) =
[\mu(r',s',\tau)]^{2}$, and where $(r',s')$ denotes any point on the
integration path.  The integration path is any segmentally smooth path
which lies entirely in $\bold{D}(\tau,z_{0})$ and which has $(a,b)$
and $(r,s)$ as initial and final points, respectively.  The integral
in Eq.\ (2.28) is defined in the sense of Lebesgue.

\item
For any given $(r,s,z_{0})$ such that $-1 \le r < z_{0} < s \le 1$,
$\bold{\xi}(r,s,\tau,z_{0})$ is a continuous function of $\tau$
throughout $\bold{D}(r,s,z_{0})$, as defined by Eq.\ (2.22), and is
holomorphic throughout $C-[r,s]$.

\item
The following relation holds for all $-1 < z_{0} < 1$:
\begin{equation}
\bold{\xi}_{1}(\infty,z_{0}) = 0
\end{equation}
where, for all $\tau \in \bold{D}(-1,1,z_{0})$,
\begin{equation}
\bold{\xi}_{1}(\tau,z_{0}) := \bold{\xi}(-1,1,\tau,z_{0}) .
\end{equation}

\item
Let ${\cal C}(\epsilon,c)$ denote any positively oriented circle in
the complex plane with radius $\epsilon$, with center $c$, and with
the points $c \pm \epsilon$ deleted.  Then, for all non-negative
integers $n$,
\begin{equation}
\int_{{\cal C}(\epsilon,c)} d\tau
(\tau-c)^{n} \bold{\xi}_{1}(\tau,z_{0}) \rightarrow 0
\mbox{ as  $\epsilon \rightarrow 0$ when $c = -1,1$, and $z_{0}$}.
\end{equation}
\end{list}
\end{description}
Note that the above integral exists since $\bold{\xi}_{1}(\tau,z_{0})$
is continuous and bounded on ${\cal C}(\epsilon,c)$.  This follows
from the continuity [condition (ii)] of $\bold{\xi}_{1}(\tau,z_{0})$
throughout $\bold{D}(-1,1,z_{0})$.

\begin{description}
\item[Notes on condition (i) in the above definition:]
When applying the integral equation (2.28), it is useful to know that
$*dr=dr$, $*ds=-ds$, $**=1$, and that for both $r < \sigma < z_{0}$
and $z_{0} < \sigma < s$,
\begin{equation}
\left( \frac{\sigma^{\pm}-s}{\sigma^{\pm}-r} \right)^{\frac{1}{2}}
= \pm i \sqrt{\frac{s-\sigma}{\sigma-r}} ,
\end{equation}
while for both $\sigma < r$ and $\sigma > s$,
\begin{equation}
\left( \frac{\sigma-s}{\sigma-r} \right)^{\frac{1}{2}}
= \sqrt{\frac{s-\sigma}{\sigma-r}} ,
\end{equation}
The existence of the integral on any segmentally smooth path in
$\bold{D}(\tau,z_{0})$ is manifest.  Also, the independence of the
integral on the choice of the integration path which joins $(a,b)$ to
$(r,s)$ requires only moderate effort to prove with the aid of Eq.\
(2.13).  The only parts of the proof which may not be obvious are
those for which the integration paths meet the dashed lines depicted
in Figs.\ 3 and 4, and the reason is that $\mu(r,s,\tau)$ vanishes on
these dashed lines.  This problem can be handled most easily by
introducing appropriate new charts for effecting the integrations in
neighborhoods of the dashed lines.  For example, in a neighborhood of
the dashed line in the domain $\bold{D}(\sigma^{\pm},z_{0})$ in Fig.\
3, one can let
\begin{eqnarray*}
\lefteqn{ \left( \frac{\sigma^{\pm}-z'-\rho' \; *}
{\sigma^{\pm}-z'+\rho' \; *} \right)^{\frac{1}{2}}
d\bold{\psi}(r',s') = } \\
& & \pm \left[ \frac{q}{\sqrt{\sigma-r(u')}} du' \psi_{u'}(u',v')
-2\sqrt{\sigma-r(u')} dq \bold{\psi}_{s'}(r(u'),\sigma+q^{2}) \right]
\end{eqnarray*}
where we used Eq.\ (2.33) and then introduced the new coordinates
$u'$ and $q = \sqrt{s'-\sigma}$, with $v'$ defined by $s(v')=\sigma
+q^{2}$.  Observe that $-1 \le r(u') < z_{0}$ and $q=0$ on the dashed
line which borders $\bold{D}(\sigma^{\pm},z_{0})$ in Fig.\ 3.  We
leave further details concerning the proof of path independence of the
integral (2.28) for the interested reader.  Also, we leave it to the
reader to establish that Eq.\ (2.28) implies that
$\bold{\xi}(\tau,z_{0})$ is continuous throughout its domain
$\bold{D}(\tau,z_{0})$, and that $d\bold{\xi}(r,s,\tau,z_{0})$ exists
and satisfies Eq.\ (2.15) at all points $(r,s)$ at which
$\mu(r,s,\tau) \ne 0$.

\item[Notes on conditions (iv) in the definition of the special
gauge:]
The combined conditions (ii), (iii) and (iv) enable us to regard
$\bold{\xi}_{1}(\tau,z_{0})$ as if it is the usual complex representation
of a two-dimensional electrostatic field whose sources all lie on the
real axis interval $[-1,1]$.  The conditions (ii) inform us that there
are continous summable (Lebesgue integrable) source distributions on
the open intervals $]-1,z_{0}[$ and $]z_{0},1[$ but that there are no
other sources on these open intervals.  The conditions (iv), as given
by (2.32), reassure us that there are no point multipole sources (of
any order, whatsoever) at the endpoints of these open intervals.  In
other words, there are no sources which produce poles or isolated
essential singularities of $\bold{\xi}_{1}(\tau,z_{0})$ at the points
$\tau = -1,1,z_{0}$.  So, in effect, the only sources for
$\bold{\xi}_{1}(\tau,z_{0})$ are the continuous summable line
distributions on $]-1,z_{0}[$ and $]z_{0},1[$.

Nevertheless, the limits of $\bold{\xi}_{1}(\tau,z_{0})$ as $\tau
\rightarrow -1,1$ or $z_{0}$ do not generally exist.  This is not
surprising if we employ the electrostatic analogue.  However, a few
explanations are in order.  Let us start with $\tau = \pm 1$.

When the initial data functions $\bold{\psi}_{3}(r)$ and
$\bold{\psi}_{2}(s)$ have continuous first derivatives at $r = -1$ and
$s = 1$, respectively, then the limits of $\bold{\xi}_{1}(\tau,z_{0})$
as $\tau \rightarrow -1$ and as $\tau \rightarrow 1$ turn out to
exist.  For example, these limits exist for the Kasner metric
$\bold{\xi}$-potentials given by Eqs.\ (2.20) and (2.21).  However,
they do not generally exist.  For example, they do not exist for any
\CGPWP.\cite{39}

As regards the singularity at $\tau = z_{0}$, we shall see that it is
always logarithmic when it exists, that it arises from a finite step
discontinuity of the source distribution at $\tau = z_{0}$, and that
it does not exist if and only if $\bold{\psi}(r,s)$ has a limit as
$(r,s) \rightarrow (z_{0},z_{0})$.
\end{description}

\subsection{The Function $\Xi$ and a Theorem}
We shall next give some simple results which revolve around the fact
that the difference $\bold{\xi}(r,s,\tau,z_{0}) -
\bold{\xi}_{1}(\tau,z_{0})$ is independent of $z_{0}$.
\begin{description}
\item[Definition:]
For any
\begin{equation}
\tau \in (C-[-1,1]) \cup (-1,1)^{+} \cup (-1,1)^{-} \; ,
\end{equation}
let
\begin{equation}
\begin{array}{rcl}
\bold{D}(\tau) & := & \DIV \mbox{ if } \tau \in C-[-1,1] \; , \\
\bold{D}(\sigma^{\pm}) & := & \{(r,s) \in \DIV : r \le \sigma \le s \}
\mbox{ if } -1 < \sigma < 1 \; .
\end{array}
\end{equation}
The domain $\bold{D}(\sigma^{\pm})$ is depicted in Fig.\ 5.

\item[Definition:]
For given $\bold{\psi}$, let $\Xi$ denote the function whose domain is
\begin{equation}
\dom \Xi := \{ (r,s,\tau): \tau \mbox{ is in the set (2.35) and }
(r,s) \in \bold{D}(\tau) \}
\end{equation}
and whose values are
\begin{equation}
\Xi(r,s,\tau) := \int_{(-1,1)}^{(r,s)}
\left( \frac{\tau - z' - \rho' \; *}{\tau - z' + \rho' \; *}
\right)^{\frac{1}{2}} d\bold{\psi}(r',s')
\end{equation}
where $(r',s')$ is any point on the integration path which joins
$(-1,1)$ to $(r,s)$ and which lies in $\bold{D}(\tau)$. Statements
similar to those made for $\bold{\xi}$ [after Eq.\ (2.28) and after
Eq.\ (2.34)] are also applicable to $\Xi$, though allowances must be
made for the fact that the domains of $\bold{\xi}$ and $\Xi$ differ.

\item[Theorem:] \
\begin{list}{(\theenumiii)}{\usecounter{enumiii}}
\item
If $(r,s,\tau) \in \dom \Xi$, then
\begin{equation}
\bold{\xi}(r,s,\tau,z_{0}) - \bold{\xi}_{1}(\tau,z_{0}) =
\Xi(r,s,\tau)
\end{equation}
for all $z_{0}$ such that $r < z_{0} < s$, and such that $z_{0} \ne
\sigma$ when $-1 < \sigma < 1$ and $\tau = \sigma^{\pm}$.  In
particular, note the choice $z_{0}=z=(s+r)/2$.

\item
If $(r,s) \in \DIV$, then
\begin{equation}
\bold{\xi}(r,s,\infty,z_{0}) = \bold{\psi}(r,s)
\end{equation}
for all $z_{0}$ such that $r < z_{0} < s$.  Moreover,
\begin{equation}
\bold{\xi}(r,s,\infty,z) = \Xi(r,s,\infty) = \bold{\psi}(r,s) .
\end{equation}
\end{list}
\end{description}

{\em Proof:} The proof is simple and uses Eqs.\ (2.28), (2.30), (2.38)
and our convention $\bold{\psi}(-1,1) := 0$.  The depiction of the
various domains in Figs.\ 2 to 5 may be helpful.

\subsection{The Spectral Function $\lambda$}

\begin{description}
\item[Theorem:]
Suppose $\tau = \sigma$ and $-1 < \sigma < z_{0}$ or $z_{0} < \sigma <
1$.  Then
\begin{subequations}
\begin{equation}
\lambda(\sigma,z_{0}) := \bold{\xi}(\sigma,s,\sigma,z_{0}) \mbox{ if }
-1 < \sigma < z_{0}
\end{equation}
and
\begin{equation}
\lambda(\sigma,z_{0}) := \bold{\xi}(r,\sigma,\sigma,z_{0}) \mbox{ if }
z_{0} < \sigma < 1
\end{equation}
\end{subequations}
are independent of $s$ over the range $z_{0} < s \le 1$ and of $r$
over the range $-1 \le r < z_{0}$, respectively.
\end{description}

{\em Proof:}  The above theorem follows easily from Eq.\ (2.28) and
the relations $r=z-\rho$ and $s=z+\rho$.

Note that the theorem contains a definition of the function $\lambda$
which appeared in Eqs.\ (1.23) and (1.24).  Those equations will be
derived in Sec.\ III.  As a point of interest for those who are
acquainted with previous work on KC transformations, Eqs.\ (2.42a) and
(2.42b) are generalizations of the axis relation for the special case
of the vacuum Weyl metrics.  As we have already stated in Sec.\ I and
as we shall show in Sec.\ IV, $\bold{\psi}$ has a continuous extension
to $\bold{D}_{IV}$ (see Eq.\ (2.1a)) if and only if
$\bold{\xi}(r,s,\tau,z_{0})$ is independent of $z_{0}$.
In that case, $\lambda(\sigma,z_{0})$ is
independent of $z_{0}$ and it will be shown that
\begin{equation}
\lambda(\sigma) := \lambda(\sigma,z_{0}) = \bold{\psi}(\sigma,\sigma) .
\end{equation}
The above Eq.\ (2.43) is the axis relation for the special case of the
vacuum Weyl metrics.

The definition of $\lambda$ which was given above was suggested by
Figs.\ 3 and 4, which depict the $(r,s)$-plane for fixed
$(\sigma,z_{0})$.  The set of points on which
$\bold{\xi}(r,s,\sigma,z_{0})$ and
$\bold{\xi}(r,s,\sigma^{\pm},z_{0})$ take on the values
$\lambda(\sigma,z_{0})$ are represented by the dashed lines in those
figures.

An equivalent definition can be based on a picture of the
$(\tau,z_{0})$-space for fixed $(r,s)$.  Specifically, given that
$(r,s) \in \DIV$,
\begin{subequations}
\begin{equation}
\lambda(r,z_{0}) := \bold{\xi}(r,s,r,z_{0})
\end{equation}
and
\begin{equation}
\lambda(s,z_{0}) := \bold{\xi}(r,s,s,z_{0})
\end{equation}
\end{subequations}
for all $z_{0}$ such that $r < z_{0} < s$.  Thus, $\lambda(r,z_{0})$
and $\lambda(s,z_{0})$ are the values of $\bold{\xi}(r,s,\tau,z_{0})$
at the end point $r$ of the pair of lips $(r,z_{0})^{\pm}$ and at the
end point $s$ of the pair of lips $(z_{0},s)^{\pm}$, respectively.

%
\setcounter{equation}{0}

\section{The Spectral Function $\xi$ as a Solution of a Hilbert Problem}

\subsection{A Preliminary Theorem on Cauchy Integrals}

\begin{description}
\item[Definition of $\Gamma^{+}$ and $\Gamma^{-}$ for a given closed
contour $\Gamma$:]  Suppose $\Gamma$ denotes a simple, segmentally
smooth, positively oriented closed contour in the complex plane C.  Then
$\Gamma^{+}$ and $\Gamma^{-}$ will denote those disjoint open subsets of
C which are bounded and unbounded, respectively, which have $\Gamma$ as
their common boundary, and which satisfy $C = \Gamma \cup \Gamma^{+}
\cup \Gamma^{-}$.  This is depicted by Fig.\ 6.

\item[Definitions of $[a,b[$ and \mbox{$]a,b]$}:]
Suppose $a$, $b$ and $d$ are real numbers such that $a < d < b$.  Then
$$[a,d[ := \{ \sigma: a \le \sigma < d \}$$
and
$$]d,b] := \{ \sigma: d < \sigma \le b \} .$$

\item[Theorem:]  Consider the function $F(\tau)$ which is defined by the
Cauchy integral of Eq.\ (2.8), where the summable function $f(\sigma)$
in the integrand has $]a,b[$ as its domain and obeys a H\"{o}lder
condition on any closed subinterval of $]a,b[$.  Let $\Gamma$ denote any
simple, segmentally smooth, positively oriented closed contour in C such
that
\begin{equation}
d \in \Gamma \; , \; [a,d[ \subset \Gamma^{+} \mbox{ and }
]d,b] \subset \Gamma^{-} \; .
\end{equation}
Then, for any non-negative integer $n$,
\begin{equation}
\int_{\Gamma-\{d\}} d\tau (\tau-a)^{n} F(\tau)
= -2i \int_{a}^{d} d\sigma (\sigma-a)^{n} f(\sigma) .
\end{equation}
If, instead, $\Gamma$ is chosen so that
\begin{equation}
d \in \Gamma \; , \; ]d,b] \subset \Gamma^{+} \mbox{ and }
[a,d[ \subset \Gamma^{-} \; .
\end{equation}
then Eq.\ (3.2) still holds provided one makes the replacements $i
\rightarrow -i$ and $a \rightarrow b$.  NOTE:  The integral on the left
side of Eq.\ (3.2) exists since $F(\tau)$ is continuous and bounded on
$\Gamma-\{d\}$.  This follows from the Plemelj theorem in Sec.\ IIC.
\end{description}

Proof:  Let $a'$ and $b'$ be any real points such that $a < a' < d < b'
< b$.  Then
\begin{eqnarray}
\int_{\Gamma-\{d\}} d\tau (\tau-a)^{n} F(\tau)
& = & \frac{1}{\pi} \int_{\Gamma-\{d\}} d\tau
\int_{a}^{a'} d\sigma \frac{(\tau-a)^{n} f(\sigma)}{(\sigma-\tau)}
\nonumber \\
& & + \frac{1}{\pi} \int_{\Gamma-\{d\}} d\tau
\int_{b'}^{b} d\sigma \frac{(\tau-a)^{n} f(\sigma)}{(\sigma-\tau)} \\
& & + \int_{\Gamma-\{d\}} d\tau (\tau-a)^{n} F(a',b',\tau) ,
\nonumber
\end{eqnarray}
where
\begin{equation}
F(a',b',\tau) := \frac{1}{\pi} \int_{a'}^{b'} d\sigma
\frac{f(\sigma)}{\sigma-\tau} .
\end{equation}
The orders of integration in the first two iterated integrals on the
right side of Eq.\ (3.4) can be interchanged. Upon interchanging the
orders of integration and completing the integrations over $\tau$, one
obtains
\begin{eqnarray}
\int_{\Gamma-\{d\}} d\tau (\tau-a)^{n} F(\tau)
& = & -2i \int_{a}^{a'} d\sigma (\sigma-a)^{n} f(\sigma) \\
& & + \frac{1}{\pi} \int_{\Gamma-\{d\}} d\tau (\tau-a)^{n}
F(a',b',\tau) . \nonumber
\end{eqnarray}
{}From the Plemelj theorem in Sec.\ IIC, $F(a',b',\tau)$ is a holomorphic
function of $\tau$ throughout $C-[a',b']$ and is a continuous function
of $\tau$ throughout the union of $C-[a',b']$, $(a',b')^{+}$ and
$(a',b')^{-}$ such that
\begin{equation}
F(a',b',\sigma^{+})-F(a',b',\sigma^{-}) = 2i f(\sigma)
\end{equation}
for all $a' < \sigma < b'$.  Another well known theorem of Plemelj tells
us that the singularity of $F(a',b',\tau)$ at $\tau=a'$ is merely
logarithmic.  Therefore, the path of integration $\Gamma-\{d\}$ in the
second term on the right side of Eq.\ (3.6) can be deformed (without
changing the value of the integral) until it becomes
$$\frac{1}{\pi} \int_{d}^{a'} d\sigma (\sigma-a)^{n}
[F(a',b',\sigma^{+})-F(a',b',\sigma^{-})] .$$
So, from Eqs.\ (3.6) and (3.7), we obtain our final conclusion (3.2).
The proof for the alternative case when the conditions (3.3) hold is
similar.

\begin{description}
\item[Corollary:]  Let $c=a$ or $b$, and let ${\cal C}(\epsilon,c)$
denote the positively oriented circle with radius $\epsilon$, with
center $c$, and with the points $c \pm \epsilon$ deleted.  Then, for any
non-negative integer $n$, the function $F(\tau)$ defined by the Cauchy
integral in Eq.\ (2.8) satisfies:
\begin{equation}
\int_{{\cal C}(\epsilon,c)} d\tau (\tau-a)^{n} F(\tau)
\rightarrow 0 \mbox{ as } \epsilon \rightarrow 0 .
\end{equation}
\end{description}

\subsection{A Hilbert Problem on $[-1,1]$}

In Sec.\ III, as before, our only premises concerning the initial data
functions will be those given in Sec.\ IIA.  Also, we shall continue to
assume that $\bold{\xi}$ exists until Sec.\ IIIC.

We now apply the integral equation (2.28) to the straight line paths
$$(r,\sigma) \rightarrow (r,s) \mbox{ where } -1 \le r < z_{0} < \sigma
\le s \le 1$$
and
$$(\sigma,s) \rightarrow (r,s) \mbox{ where } -1 \le r \le \sigma <
z_{0} < s \le 1$$
in the domains $\bold{D}(\sigma^{\pm},z_{0})$ which are depicted in
Figs.\ 3 and 4, respectively.  With the aid of Eqs.\ (2.33), (2.42a) and
(2.42b), we obtain
\begin{subequations}
\begin{equation}
\bold{\xi}(r,s,\sigma^{\pm},z_{0}) = \lambda(\sigma,z_{0})
\mp i \int_{\sigma}^{s} ds' \sqrt{\frac{\sigma-r}{s'-\sigma}}
\bold{\psi}_{s'}(r,s')
\end{equation}
for $z_{0} < \sigma < s$, and
\begin{equation}
\bold{\xi}(r,s,\sigma^{\pm},z_{0}) = \lambda(\sigma,z_{0})
\pm i \int_{\sigma}^{r} dr' \sqrt{\frac{s-\sigma}{\sigma-r'}}
\bold{\psi}_{r'}(r',s)
\end{equation}
\end{subequations}
for $r < \sigma < z_{0}$.  The above Eqs.\ (3.9a) and (3.9b) will be
used both here and in Sec.\ IIIE.

Here, we apply Eqs.\ (3.9a) and (3.9b) when $(r,s)=(-1,1)$.  From the
definitions (1.18) and (1.21) of $\G_{j}(\sigma)$ in terms of the
initial data functions, from the definition (1.22) of
$\G(\sigma,z_{0})$, and from the definition (2.31) of
$\bold{\xi}_{1}(\tau,z_{0})$,
\begin{equation}
\bold{\xi}_{1}(\sigma^{\pm},z_{0}) = \lambda(\sigma,z_{0})
\mp i \G(\sigma,z_{0})
\end{equation}
or, equivalently,
\begin{equation}
\bold{\xi}_{1}(\sigma^{+},z_{0}) - \bold{\xi}_{1}(\sigma^{-},z_{0})
= - 2i \G(\sigma,z_{0})
\end{equation}
and
\begin{equation}
\bold{\xi}_{1}(\sigma^{+},z_{0}) + \bold{\xi}_{1}(\sigma^{-},z_{0})
= 2 \lambda(\sigma,z_{0}) .
\end{equation}
The task of finding $\bold{\xi}_{1}(\tau,z_{0})$ so that Eq.\ (3.11) is
satisfied is an example of a simple kind of Hilbert problem.

\begin{description}

\item[A Hilbert Problem on \mbox{$[-1,1]$}:]  Let the function
$\G(\sigma,z_{0})$ be given in terms of the initial data functions.
We seek a function $\bold{\xi}_{1}$ with the domain
$$\{(\tau,z_{0}): -1 < z_{0} < 1 \mbox{ and } \tau \in
\bold{D}(-1,1,z_{0}) \}$$
such that conditions (ii), (iii) and (iv) in Sec.\ IIE are satisfied
by $\bold{\xi}_{1}$ and such that Eq.\ (3.11) holds.

\item[Theorem:]  There exists exactly one solution
\begin{equation}
\bold{\xi}_{1}(\tau,z_{0}) = - \frac{1}{\pi}
\int_{-1}^{1} d\sigma \frac{\G(\sigma,z_{0})}{\sigma-\tau}
\end{equation}
of the preceding Hilbert problem.
\end{description}

Proof:  Employ the two theorems in Sec.\ IIC and the corollary [Eq.\
(3.8)] in Sec. IIIA to prove that Eq.\ (3.13) is a solution of the
Hilbert problem.  As regards uniqueness, let
$$\Delta := \bold{\xi}_{1}' - \bold{\xi}_{1}$$
for any given solutions $\bold{\xi}_{1}$ and $\bold{\xi}_{1}'$ of the
Hilbert problem.  Conditions (ii) and (iii) in Sec.\ IIE, taken together
with Eq.\ (3.11), imply that $\Delta(\tau,z_{0})$ is a continuous
function of $\tau$ throughout
$$(C-[-1,1]) \cup (-1,z_{0})^{+} \cup (-1,z_{0})^{-}
\cup (z_{0},1)^{+} \cup (z_{0},1)^{-},$$
that $\Delta(\tau,z_{0})$ is holomorphic throughout $C-[-1,1]$, that
$\Delta(\infty,z_{0}) = 0$, and that $\Delta(\sigma^{+},z_{0}) =
\Delta(\sigma^{-},z_{0})$ for all $-1 < \sigma < z_{0}$ and $z_{0} <
\sigma < 1$.  A well known theorem of Riemann then implies that
$\Delta(\tau,z_{0})$ has a holomorphic continuation to the domain
$C-\{-1,1,z_{0}\}$.  However, condition (iv) [Eq.\ (2.32)] in Sec.\ IIE
tells us that there are no isolated singularities at $\tau=-1,1,z_{0}$.
So, from Liouville's theorem and the equation $\Delta(\infty,z_{0}) =
0$, it follows that $\Delta := \bold{\xi}_{1}' - \bold{\xi}_{1}$ is
identically zero.

\begin{description}
\item[Corollary:]  The solution for $\lambda(\sigma,z_{0})$ is given by
Eq.\ (1.23).  Moreover, $\lambda(\sigma,z_{0})$ obeys a H\"{o}lder
condition of index $\frac{1}{2}$ on every closed subinterval of the open
intervals $-1 < \sigma < z_{0}$ and $z_{0} < \sigma < 1$.  Finally,
\begin{equation}
\bold{\xi}(r,s,\sigma^{+},z_{0}) + \bold{\xi}(r,s,\sigma^{-},z_{0})
= 2 \lambda(\sigma,z_{0})
\end{equation}
for all $(r,s)$ in $\bold{D}(\sigma^{\pm},z_{0})$.
\end{description}

Proof:  Use the two theorems in Sec.\ IIC as well as Eqs.\ (3.12) and
(3.13).  Equation (3.14) is implied directly by Eqs.\ (3.9a) and
(3.9b).

\subsection{The Existence and Uniqueness of $\xi$}

\begin{description}
\item[Theorem:]  For given $\bold{\psi}$, a spectral potential
$\bold{\xi}$ in the special gauge exists.
\end{description}

Proof:  In this proof, $\bold{\xi}_{1}(\tau,z_{0}) :=
\bold{\xi}(-1,1,\tau,z_{0})$ will be defined by Eq.\ (3.13) for all
$\tau$ in $\bold{D}(-1,1,z_{0})$.  [See Eq.\ (2.22).]

The first step is to introduce the function $\bold{\xi}^{(1)}$ whose
domain is
\begin{eqnarray}
\dom{\bold{\xi}^{(1)}} & := & \{ (r,s,\tau,z_{0}):
(\tau,z_{0}) \in \dom{\bold{\xi}_{1}} \nonumber \\
& & \mbox{ and }
(r,s) \in \bold{D}(\tau,z_{0}) \} \nonumber \\
& = & \{ (r,s,\tau,z_{0}): -1 \le r < z_{0} < s \le 1 \nonumber \\
& & \mbox{ and }
\tau \in \bold{D}(-1,1,z_{0}) \}
\end{eqnarray}
and whose values are
\begin{equation}
\bold{\xi}^{(1)}(r,s,\tau,z_{0}) := \bold{\xi}_{1}(\tau,z_{0})
+ \Xi(r,s,\tau) ,
\end{equation}
where $\bold{D}(\tau,z_{0})$ is defined by Eq.\ (2.24a) and depicted by
Fig.\ 2 when $\tau \in C-[-1,1]$, where $\bold{D}(\sigma^{\pm},z_{0})$
is defined by Eq.\ (2.24d) and depicted by Fig.\ 3 when $z_{0} < \sigma
< 1$, where $\bold{D}(\sigma^{\pm},z_{0})$ is defined by Eq.\ (2.24f)
and depicted by Fig.\ 4 when $-1 < \sigma < z_{0}$, and where $\Xi$ is
defined in Sec.\ IIF.  By using alternative paths of integration in the
defining equation (2.38) for $\Xi$, we obtain
\begin{equation}
\Xi(r,s,\tau) = \int_{-1}^{r} dr' \left( \frac{\tau-1}{\tau-r'}
\right)^{\frac{1}{2}} \dot{\bold{\psi}}_{3}(r')
+ \int_{1}^{s} ds' \left( \frac{\tau-r}{\tau-s'} \right)^{\frac{1}{2}}
\bold{\psi}_{s'}(r,s')
\end{equation}
and
\begin{equation}
\Xi(r,s,\tau) = \int_{1}^{s} ds' \left( \frac{\tau+1}{\tau-s'}
\right)^{\frac{1}{2}} \dot{\bold{\psi}}_{2}(s')
+ \int_{-1}^{r} dr' \left( \frac{\tau-s}{\tau-r'} \right)^{\frac{1}{2}}
\bold{\psi}_{r'}(r',s) ,
\end{equation}
where recall that $\bold{\psi}_{3}(r) := \bold{\psi}(r,1)$ and
$\bold{\psi}_{2}(s) := \bold{\psi}(-1,s)$.  From the above Eq.\ (3.16),
we obtain with the aid of Eqs.\ (1.18), (1.21), (1.22), (2.33), (2.34)
and (3.11):
\begin{equation}
\bold{\xi}^{(1)}(r,s,\sigma^{+},z_{0}) -
\bold{\xi}^{(1)}(r,s,\sigma^{-},z_{0}) = 0 \mbox{ for } -1 < \sigma \le r
\end{equation}
by employing Eq.\ (3.17), and
\begin{equation}
\bold{\xi}^{(1)}(r,s,\sigma^{+},z_{0}) -
\bold{\xi}^{(1)}(r,s,\sigma^{-},z_{0}) = 0 \mbox{ for } s \le \sigma < 1
\end{equation}
by employing Eq.\ (3.18).

The second phase of the proof uses Eqs.\ (3.19) and (3.20) to define a
new function $\bold{\xi}^{(2)}$ whose domain is [compare with Eq.\
(3.15)]
\begin{eqnarray}
\dom{\bold{\xi}^{(2)}} & := & \{ (r,s,\tau,z_{0}): -1 \le r < z_{0} < s \le
1 \nonumber \\
& & \mbox{ and } \tau \in [\bold{D}(r,s,z_{0})-\{-1,1\} \}
\end{eqnarray}
and whose values are
\begin{eqnarray}
\bold{\xi}^{(2)}(r,s,\tau,z_{0}) & := & \bold{\xi}^{(1)}(r,s,\tau,z_{0})
\nonumber \\
& \mbox{when} & \tau \in C-[-1,1] , \\
& \mbox{or when} & \tau = \sigma^{\pm} \mbox{ and } r < \sigma < s ,
\nonumber
\end{eqnarray}
and, using Eqs.\ (3.19) and (3.20),
\begin{eqnarray}
\bold{\xi}^{(2)}(r,s,\sigma,z_{0}) & := &
\bold{\xi}^{(1)}(r,s,\sigma^{\pm},z_{0}) \nonumber \\
& \mbox{when} & -1 < \sigma \le r \mbox{ or } s \le \sigma < 1 .
\end{eqnarray}
{}From its definition by Eqs.\ (3.15) and (3.16),
$\bold{\xi}^{(1)}(r,s,\tau,z_{0})$ is (for fixed $r,s,z_{0}$)
holomorphic throughout $C-[-1,1]$ and continuous throughout
$$(C-[-1,1]) \cup (-1,z_{0})^{+} \cup (-1,z_{0})^{-}
\cup (z_{0},1)^{+} \cup (z_{0},1)^{-}.$$
Therefore, from Eq.\ (3.21) to Eq.\ (3.23) and a well known theorem of
Riemann, $\bold{\xi}^{(2)}(r,s,\tau,z_{0})$ is holomorphic throughout
\begin{equation}
C-([r,s] \cup \{-1,1\})
\end{equation}
and is a continuous function of $\tau$ throughout
$$C-(]r,s[ \cup \{-1,1\}) \cup (r,z_{0})^{+} \cup (r,z_{0})^{-}
\cup (z_{0},1)^{+} \cup (z_{0},1)^{-} .$$
Note that the above set includes the point $\tau=r$ when $r > -1$ and
the point $\tau=s$ when $s < 1$.

The final phase of the proof considers the two isolated points
$$\tau=-1 \mbox{ when } r > -1 \; , \; \tau=1 \mbox{ when } s < 1$$
which are excluded from the domain of holomorphy (3.24).  Upon applying
the corollary [Eq.\ (3.8)] in Sec.\ IIIA to the expression for
$\bold{\xi}_{1}$ in Eq.\ (3.13) and upon using Eqs.\ (3.16), (3.17),
(3.18), (3.22) and (3.23), we see that $\bold{\xi}^{(2)}(r,s,\tau,z_{0})$
has no pole and no isolated essential singularity at either of these
points.

Now let $\bold{\xi}$ denote that function whose domain is [compare with
$\dom{\bold{\xi}^{(2)}}$ in Eq.\ (3.21)] given by Eq.\ (2.25) and whose
values are
\begin{equation}
\bold{\xi}(r,s,\tau,z_{0}) := \bold{\xi}^{(2)}(r,s,\tau,z_{0})
\mbox{ when } (r,s,\tau,z_{0}) \in \dom{\bold{\xi}^{(2)}}
\end{equation}
and
\begin{subequations}
\begin{eqnarray}
\bold{\xi}(r,s,-1,z_{0}) & := & \lim_{\tau \rightarrow -1}
\bold{\xi}^{(2)}(r,s,\tau,z_{0}) \mbox{ when } r > -1 , \\
\bold{\xi}(r,s,1,z_{0}) & := & \lim_{\tau \rightarrow 1}
\bold{\xi}^{(2)}(r,s,\tau,z_{0}) \mbox{ when } s < 1 .
\end{eqnarray}
\end{subequations}
The above limits exist, of course.  We leave it to the enterprising
reader to verify that the function $\bold{\xi}$ which has just been
defined satisfies all of the requirements specified in the definition of
the special gauge in Sec.\ IIE.

\begin{description}
\item[Theorem:]  For given $\bold{\psi}$, the spectral potential
$\bold{\xi}$ in the special gauge is unique.
\end{description}

Proof:  We proved that $\bold{\xi}_{1}$ is unique in Sec.\ IIIB.  The
uniqueness of $\bold{\xi}$ then follows from Eq.\ (2.39) in the theorem
of Sec.\ IIF.

\subsection{The Spectral Potential $\phi$}

\begin{description}
\item[Definition:] Let
\begin{equation}
\bold{\phi} := \bold{\xi}/\mu .
\end{equation}
The domain of $\bold{\phi}$ is the domain of $\bold{\xi}$ minus those
points $(r,s,\tau,z_{0})$ at which $\mu(r,s,\tau)=0$, i.e., at which
$\tau=r$ or $\tau=s$.  From Eqs.\ (2.12) and (2.15), one derives:
\begin{equation}
d * (\rho d\bold{\phi}) = 0 .
\end{equation}
So $\bold{\phi}$ satisfies the same hyperbolic equation (2.2) as
$\bold{\psi}$.
\end{description}

\subsection{A Hilbert Problem on $[r,s]$}

With the aid of Eqs.\ (2.6) and (3.27), one sees that Eqs.\ (3.9a) and
(3.9b) imply that
\begin{equation}
\bold{\phi}(r,s,\sigma^{+},z_{0}) - \bold{\phi}(r,s,\sigma^{-},z_{0})
= \frac{-2i \lambda(\sigma,z_{0})}{\sqrt{(\sigma-r)(s-\sigma)}} .
\end{equation}
This equation is the basis for another simple Hilbert problem.  Both in
the statement of the Hilbert problem and in the derivation of its
solution, we shall grant the relation
\begin{equation}
\lambda(\sigma,z_{0}) = \Lambda(\sigma,z_{0})
+ \frac{1}{\pi} [\G_{3}(z_{0})-\G_{2}(z_{0})]
\ln{|\sigma-z_{0}|}
\end{equation}
where $\Lambda(\sigma,z_{0})$ is defined throughout the open
interval $-1 < \sigma < 1$ and obeys a H\"{o}lder condition of index
$\frac{1}{2}$ on every closed subinterval of $]-1,1[$.  Equation (3.30)
will be proven in Sec.\ IV.  From Eq.\ (3.30), it follows that the right
side of Eq.\ (3.29) obeys a H\"{o}lder condition of index $\frac{1}{2}$
on every closed subinterval of the open intervals $r < \sigma < z_{0}$
and $z_{0} < \sigma < s$.  It also follows that the right side of Eq.\
(3.29) is summable on $[r,s]$ if $-1 < r < z_{0} < s < 1$.

\begin{description}
\item[A Hilbert Problem on \mbox{$[r,s]$}:]  At present, assume that $-1 <
r < z_{0} < s < 1$.  Let $\lambda(\sigma,z_{0})$ be given in terms of the
initial data functions by Eqs.\ (1.18) to (1.23).  Then, for fixed
$(r,s,z_{0})$, we seek $\bold{\phi}(r,s,\tau,z_{0})$ such that it is a
holomorphic function of $\tau$ throughout $C-[r,s]$, it is a continuous
function of $\tau$ throughout $\bold{D}(r,s,z_{0})-\{r,s\}$, it
satisfies Eq.\ (3.29) and has the value $0$ at $\tau=\infty$, and it
satisfies the condition that
$$\mu(r,s,\tau) |\tau-z_{0}|^{\epsilon} \bold{\phi}(r,s,\tau,z_{0})$$
(for arbitrary $\epsilon > 0$) remain bounded as $\tau \rightarrow r$,
as $\tau \rightarrow s$, and as $\tau \rightarrow z_{0}$.
\end{description}

The solution of the above Hilbert problem is well known, is unique and
is given by
\begin{equation}
\bold{\phi}(r,s,\tau,z_{0}) = - \frac{1}{\pi} \int_{r}^{s} d\sigma
\frac{\lambda(\sigma,z_{0})}{(\sigma-\tau) \sqrt{(\sigma-r)(s-\sigma)}}.
\end{equation}
Equation (3.29) can be extended to include $r=-1$ and $s=1$ by using the
fact that $\bold{\phi}(r,s,\tau,z_{0})$ is continuous at $r=-1$ when
$\tau \ne -1$ and at $s=1$ when $\tau \ne 1$:
\begin{equation}
\begin{array}{rcl}
\bold{\phi}(-1,s,\tau,z_{0}) & = & \lim_{r \rightarrow -1}
\bold{\phi}(r,s,\tau,z_{0}) \mbox{ when } \tau \ne -1 , \\
\bold{\phi}(r,1,\tau,z_{0}) & = & \lim_{s \rightarrow 1}
\bold{\phi}(r,s,\tau,z_{0}) \mbox{ when } \tau \ne 1 .
\end{array}
\end{equation}

{}From Eqs.\ (3.27) and (3.31)
\begin{equation}
\bold{\xi}(r,s,\tau,z_{0}) = - \frac{\mu(r,s,\tau)}{\pi}
\int_{r}^{s} d\sigma \frac{\lambda(\sigma,z_{0})}
{(\sigma-\tau)\sqrt{(\sigma-r)(s-\sigma)}}
\end{equation}
whereupon Eq.\ (1.24) follows from Eqs.\ (2.5) and (2.41).  The
alternative expression for $\bold{\psi}$ in Eq.\ (1.24) is obtained by
the substitution $\sigma=z+\rho \sin{\theta}$.  The same substitution
yields an alternative expression for the integral in Eq.\ (3.33).  By
setting $(r,s)=(-1,1)$ in Eq.\ (3.33), we obtain
\begin{equation}
\bold{\xi}_{1}(\tau,z_{0}) = - \frac{\mu(-1,1,\tau)}{\pi}
\int_{-1}^{1} d\sigma \frac{\lambda(\sigma,z_{0})}
{(\sigma-\tau)\sqrt{1-\sigma^{2}}} .
\end{equation}
{}From the Plemelj theorem in Sec.\ IIC and from Eqs.\ (3.11) and (2.6),
Eq.\ (3.34) yields
\begin{equation}
\G(\sigma,z_{0}) = \frac{\sqrt{1-\sigma^{2}}}{\pi}
\int_{-1}^{1} d\sigma' \frac{\lambda(\sigma',z_{0})}
{(\sigma'-\sigma)\sqrt{1-\sigma^{\prime 2}}}.
\end{equation}
The above Eq.\ (3.35) is the inverse relation of Eq.\ (1.23).

%
\setcounter{equation}{0}

\section{The Singularities of $\xi$ and $\psi$ at $\rho=0$}

\subsection{The Singularities of $\xi_{1}(\tau,z_{0})$ at $\tau=z_{0}$
and of $\lambda(\sigma,z_{0})$ at $\sigma=z_{0}$}

Observe that $\G(\sigma,z_{0})$ as defined by Eq.\ (1.22) generally has
a step discontinuity at $\sigma=z_{0}$.  Let
\begin{equation}
\tilde{\G}(\sigma,z_{0}) := \left\{
\begin{array}{ccc}
\G_{3}(\sigma)-\G_{3}(z_{0}) & \mbox{if} & \sigma \le z_{0} \\
\G_{2}(\sigma)-\G_{2}(z_{0}) & \mbox{if} & z_{0} \le \sigma.
\end{array}
\right.
\end{equation}
Then $\tilde{\G}(\sigma,z_{0})$ is continuous at $\sigma=z_{0}$ and,
from the second theorem in Sec.\ IIC and the definition of a H\"{o}lder
condition, $\tilde{\G}(\sigma,z_{0})$ obeys a H\"{o}lder condition of
index $\frac{1}{2}$ on any given closed subinterval of $-1 < \sigma < 1$.
Upon substituting from Eq.\ (4.1) into Eqs.\ (1.23) and (3.13), we obtain
\begin{equation}
\lambda(\sigma,z_{0}) = \tilde{\lambda}(\sigma,z_{0})
- \frac{1}{\pi} \G_{3}(z_{0})
\ln \left( \frac{|\sigma-z_{0}|}{\sigma+1} \right)
- \frac{1}{\pi} \G_{2}(z_{0})
\ln \left( \frac{1-\sigma}{|\sigma-z_{0}|} \right)
\end{equation}
and
\begin{equation}
\bold{\xi}_{1}(\tau,z_{0}) = \tilde{\bold{\xi}}_{1}(\tau,z_{0})
- \frac{1}{\pi} \G_{3}(z_{0}) \ln \left( \frac{\tau-z_{0}}{\tau+1} \right)
- \frac{1}{\pi} \G_{2}(z_{0}) \ln \left( \frac{\tau-1}{\tau-z_{0}} \right) ,
\end{equation}
where
\begin{equation}
\tilde{\lambda}(\sigma,z_{0}) := - \frac{1}{\pi} \int_{-1}^{1} d\sigma'
\frac{\tilde{\G}(\sigma,z_{0})}{\sigma'-\sigma}
\end{equation}
and
\begin{equation}
\tilde{\bold{\xi}}_{1}(\tau,z_{0}) := - \frac{1}{\pi} \int_{-1}^{1} d\sigma
\frac{\tilde{\G}(\sigma,z_{0})}{\sigma-\tau} .
\end{equation}
The principal values of the logarithms are understood in Eq.\ (4.2).
The cuts for the logarithms in Eq.\ (4.3) are $[-1,z_{0}]$ and $[z_{0},1]$,
respectively, and their values at $\tau=\infty$ are zero.  From the
Plemelj theorem in Sec.\ IIC, we obtain the following statements.

\begin{description}
\item[Properties of $\tilde{\lambda}(\sigma,z_{0})$ and
$\tilde{\xi}_{1}(\tau,z_{0})$ for fixed $z_{0}$:]  \

\begin{enumerate}
\item The function $\tilde{\lambda}(\sigma,z_{0})$ obeys a H\"{o}lder
condition of index $\frac{1}{2}$ on any given closed subinterval of
$-1 < \sigma < 1$.
\item The function $\tilde{\bold{\xi}}_{1}(\tau,z_{0})$ is holomorphic
throughout $C-[-1,1]$, is continuous throughout
$$(C-[-1,1]) \cup (-1,1)^{+} \cup (-1,1)^{-}$$
and satisfies the Plemelj relation
\begin{equation}
\tilde{\bold{\xi}}_{1}(\sigma^{\pm},z_{0}) = \tilde{\lambda}(\sigma,z_{0})
\mp i \tilde{\G}(\sigma,z_{0}) .
\end{equation}
\end{enumerate}
For computational purposes, it is convenient to recast the expression
(4.2) into the following form:
\begin{equation}
\lambda(\sigma,z_{0}) = \Lambda(\sigma,z_{0})
-\frac{1}{\pi} [\G_{3}(z_{0})-\G_{2}(z_{0})] \ln{|\sigma-z_{0}|}
\end{equation}
where
\begin{equation}
\Lambda(\sigma,z_{0}) := \tilde{\lambda}(\sigma,z_{0})
+\frac{1}{\pi} \G_{3}(z_{0}) \ln (1+\sigma)
-\frac{1}{\pi} \G_{2}(z_{0}) \ln (1-\sigma) .
\end{equation}
Since the logarithms of $1 \pm \sigma$ are differentiable in the open
interval $-1 < \sigma < 1$, {\em $\Lambda(\sigma,z_{0})$ obeys a
H\"{o}lder condition of index $\frac{1}{2}$ on any given closed
subinterval of $-1 < \sigma < 1$}.  We shall grant this fact below.
\end{description}

\subsection{The Singularities of the Potentials at $\rho=0$}

We shall now employ the results obtained in Sec.\ IIF.  From Eqs.\
(2.39) and (3.33) and from the fact that we may set $z_{0}=z$ in
Eq.\ (2.39),
\begin{eqnarray}
\lefteqn{\bold{\xi}(r,s,\tau,z_{0}) - \bold{\xi}_{1}(\tau,z_{0})
= \Xi(r,s,\tau)}
\nonumber \\
& = & - \bold{\xi}_{1}(\tau,z) - \frac{\mu(r,s,\tau)}{\pi} \int_{r}^{s}
d\sigma \frac{\lambda(\sigma,z)}{(\sigma-\tau)\sqrt{(\sigma-r)(s-\sigma)}} .
\end{eqnarray}
We substitute from Eq.\ (4.8) into the above Eq.\ (4.9).  The term
involving $\ln |\sigma-z|$ may be integrated explicitly.  One obtains
different forms of the integral for the two cases $\rho < |\tau-z|$ and
$\rho > |\tau-z|$.  Here, we are interested only in the case
$\rho < |\tau-z|$:
\begin{eqnarray}
\lefteqn{\bold{\xi}(r,s,\tau,z_{0})-\bold{\xi}_{1}(\tau,z_{0})
= \Xi(r,s,\tau) = \bold{\xi}_{1}(\tau,z)} \nonumber \\
& & +\frac{1}{\pi} [\G_{3}(z)-\G_{2}(z)] \left\{
\frac{\mu(r,s,\tau)}{\tau-z} \ln \left[
\frac{\tau-z+\mu(r,s,\tau)}{\tau-z} \right] - \ln \rho \right\} \nonumber \\
& & - \frac{\mu(r,s,\tau)}{\pi} \int_{r}^{s} d\sigma
\frac{\Lambda(\sigma,z)}{(\sigma-\tau)\sqrt{(\sigma-r)(s-\sigma)}}
\end{eqnarray}
for $\rho < |\tau-z|$.  Upon employing the identity
$$(\sigma-r)(s-\sigma) = \rho^{2} - (\sigma-z)^{2}$$
and introducing a new integration variable, we obtain the following
alternative expression for the integral in Eq.\ (4.10):
\begin{equation}
\int_{r}^{s} d\sigma \frac{\Lambda(\sigma,z)}
{(\sigma-\tau)\sqrt{(\sigma-r)(s-\sigma)}} =
\int_{-\frac{\pi}{2}}^{\frac{\pi}{2}} d\theta
\frac{\Lambda(z+\rho \sin \theta,z)}{z+\rho \sin \theta - \tau} .
\end{equation}
{}From Eqs.\ (2.5), (2.30) and (2.41), we further obtain
\begin{eqnarray}
\bold{\psi}(r,s) & = &
- \frac{1}{\pi} [\G_{3}(z)-\G_{2}(z)] \ln (\frac{\rho}{2})
+ \frac{1}{\pi} \int_{r}^{s} d\sigma
\frac{\Lambda(\sigma,z)}{\sqrt{(\sigma-r)(s-\sigma)}} \nonumber \\
& = & - \frac{1}{\pi} [\G_{3}(z)-\G_{2}(z)] \ln (\frac{\rho}{2})
+ \frac{1}{\pi} \int_{-\frac{\pi}{2}}^{\frac{\pi}{2}} d\theta
\Lambda(z+\rho \sin \theta,z) .
\end{eqnarray}
Thus the statement made in Sec.\ I that the expression (1.28) has a
continuous extension to
\begin{equation}
\bold{D}_{IV} := D_{IV} \cup \{ (z,z): -1 < z < 1 \}
\end{equation}
is proven.  From Eq.\ (4.10), it is also apparent that
$$\bold{\xi}(r,s,\tau,z_{0}) + \frac{1}{\pi} [\G_{3}(z)-\G_{2}(z)]
\ln (\frac{\rho}{2})$$
has a continuous extension to $\bold{D}_{IV}$ and one obtains
\begin{eqnarray}
\lefteqn{\{ \bold{\psi}(r,s) + \frac{1}{\pi} [\G_{3}-\G_{2}(z)] \ln
(\frac{\rho}{2}) \}_{\rho=0} } \nonumber \\
& = & \{ \bold{\xi}(r,s,\tau,z_{0}) - \bold{\xi}_{1}(\tau,z_{0})
+ \bold{\xi}_{1}(\tau,z)
\nonumber \\
& & + \frac{1}{\pi} [\G_{3}(z)-\G_{2}(z)] \ln (\frac{\rho}{2}) \}_{\rho=0} \\
& = & \Lambda(z,z) . \nonumber
\end{eqnarray}

As we have already indicated in Sec.\ IF, the above results are in
agreement with those already obtained by Yurtsever\cite{30}; i.e., our
results supply the same asymptotic expression for $\bold{\psi}$ as $\rho
\rightarrow 0$.  For those who wish to make a comparison, here are some
correspondences between his notations and ours.  His notations are on the
left sides of the equations:
\begin{equation}
\begin{array}{l}
U = - \ln \rho \; , \; V = -2 \psi + \ln \rho \; , \\
\alpha = \rho \; , \; \beta = z \; , \; r_{his} = -r \; , \; s_{his} = s \; .
\end{array}
\end{equation}
Of course, our expressions for $\bold{\xi}$ are completely new (since
Yurtsever made no use of the spectral potential concept) and our integral
expressions in Eqs.\ (4.9) to (4.12) are new.

A Comparison of Eq.\ (4.12) with the Kasner $\psi$-potential of Eq.\ (2.18)
shows why Yurtsever was led to regard $\bold{\psi}$ as having a
Kasner-like dependence (for fixed $z$) near $\rho=0$.

\subsection{The Potentials when the Axis ($\rho=0$) is Accessible}

\begin{description}
\item[Theorem:] \

\begin{list}{(\theenumiii)}{\usecounter{enumiii}}
\item A necessary and sufficient condition for $\bold{\psi}$ to have a
continuous extension to $\bold{D}_{IV}$ [Eq.\ (4.13)] is that
\begin{equation}
\G_{3}(\sigma) = \G_{2}(\sigma) =: \G(\sigma)
\mbox{ for all } -1 < \sigma < 1
\end{equation}
which is equivalent to the statement that $\bold{\xi}_{1}(\tau,z_{0})$
is independent of $z_{0}$ or, alternatively, to the statement that
$\lambda(\sigma,z_{0})$ is independent of $z_{0}$.

\item When the condition (4.16) holds, $\G(\sigma)$ is integrable over
$[-1,1]$ and obeys a H\"{o}lder condition of index $\frac{1}{2}$ on any
given closed subinterval of $]-1,1[$.  Also
\begin{equation}
\bold{\xi}_{1}(\tau) = - \frac{1}{\pi} \int_{-1}^{1} d\sigma
\frac{\G(\sigma)}{\sigma-\tau}
\end{equation}
and
\begin{equation}
\psi(\sigma,\sigma) = \lambda(\sigma) = - \frac{1}{\pi} \int_{-1}^{1}
d\sigma' \frac{\G(\sigma')}{\sigma'-\sigma} .
\end{equation}
Furthermore,
\begin{eqnarray}
\bold{\xi}(r,s,\tau) & = & - \frac{\mu(r,s,\tau)}{\pi} \int_{r}^{s} d\sigma
\frac{\lambda(\sigma)}{(\sigma-\tau)\sqrt{(\sigma-r)(s-\sigma)}} \nonumber \\
& = & - \frac{\mu(r,s,\tau)}{\pi} \int_{-\frac{\pi}{2}}^{\frac{\pi}{2}}
d\theta \frac{\lambda(z+\rho \sin \theta)}{z+\rho \sin \theta - \tau}
\end{eqnarray}
and
\begin{equation}
\bold{\psi}(r,s) = \frac{1}{\pi} \int_{r}^{s} d\sigma
\frac{\lambda(\sigma)}{\sqrt{(\sigma-r)(s-\sigma)}} =
\frac{1}{\pi} \int_{-\frac{\pi}{2}}^{\frac{\pi}{2}} d\theta
\lambda(z+\rho \sin \theta) .
\end{equation}
\end{list}
\end{description}

Proof:  Use Eqs. (1.22), (1.23), (3.13), (4.7), (4.10), (4.11) and (4.12).
Also, use the second theorem in Sec.\ IIC.

The question of what happens when we take partial derivatives with respect
to $r$ and $s$ of the above integrals will have to be saved for a future
paper (in which we shall adopt stronger premises than those given in
Sec.\ IIA).

%
\setcounter{equation}{0}

\section{A Broad Class of \CGPWP's}
\subsection{Premises and Colliding Wave Conditions}

For an arbitrary real member of $\Sigma_{\e}$, we do not know if
$\lambda(\sigma,z_{0})$ has asymptotic forms as $\sigma \rightarrow -1$ and
as $\sigma \rightarrow 1$.  The purpose of this section is to introduce a
two-parameter family of subsets $\Sigma(n_{3},n_{2}) \subset \Sigma_{\e}$
for which the asymptotic forms exist, are easy to deduce and are simple.
Moreover, $\Sigma(n_{3},n_{2}) \subset \Sigma_{\e}^{\CW}$ if and only if
each of the real parameters $n_{j}$ is either equal to $1$ or is in the
range $2 \le n_{j} < \infty$.  For all other cases, $\Sigma(n_{3},n_{2})$
and $\Sigma_{\e}^{\CW}$ are disjoint.

At present, there is no need to specialize to real Ernst potentials and
it will serve the objectives of later papers as well as this one to
consider any member $\E$ of $\Sigma_{\e}$ until we reach Sec.\ VF.  The
subsets $\Sigma(n_{3},n_{2}) \subset \Sigma_{\e}$ will be defined in Sec.\
VE.  Until then, we shall introduce some preliminary concepts.  We start
by replacing the premises given in Sec.\ IIA by the following ones which
imply those given in Sec.\ IIA and do not, therefore, effect the validity
of any results obtained in previous sections.
\begin{description}
\item[Premises:]
Until we reach Sec.\ VF, we shall assume that $r(u)$, $\E_{3}(u)$ and $s(v)$,
$\E_{2}(v)$ are ${\bf C}^{2}$ throughout $0 \le u < u_{0}$ and $0 \le v <
v_{0}$, respectively, with the possible exception of a finite number of
finite step discontinuities in $\ddot{\E}_{3}(u)$ and $\ddot{\E}_{2}(v)$.
Also, we continue to assume that the conditions (1.13) hold, i.e.,
$\dot{r}(u) > 0$ and $\dot{s}(v) < 0$ throughout $0 < u < u_{0}$ and
$0 < v < v_{0}$, respectively.
\end{description}

The initial data functions $r(u)$, $\E_{3}(u)$ and $s(v)$, $\E_{2}(v)$ for
which the corresponding Ernst potentials $\bold{\E}$ (with domain $\DIV$)
are members of $\Sigma_{\e}^{\CW}$ are those which satisfy the following
additional conditions.\cite{10}
\begin{description}
\item[Colliding Wave Conditions:] \
\begin{enumerate}
\item
Equations (1.14) ($\dot{r}(0)=\dot{s}(0)=0$) hold.
\item
The limits of the ratios
$$[\ddot{r}(u)-2(1-r(u)) |\dot{\E}_{3}(u)|^{2}]/\dot{r}(u)$$
and
$$[\ddot{s}(v)+2(1+s(v)) |\dot{\E}_{2}(v)|^{2}]/\dot{s}(v)$$
as $u \rightarrow 0$ and $v \rightarrow 0$ (from above), respectively, exist.
\end{enumerate}
\end{description}
The origin of the above conditions (2) will become a little clearer below.

\subsection{The Functions $\gamma$, $\gamma_{3}$ and $\gamma_{2}$}

So far, we have not discussed the metric component $g_{34}$ in the line
element (1.1).  Let
\begin{equation}
\gamma := \frac{1}{2} \ln (f g_{34}) \; , \; f := \Re \E = -g_{22} \; .
\end{equation}
In our current work, the vacuum field equations for the line element (1.1)
are all expressed in terms of the functions $\rho$, $\E$ and $\gamma$.
In some previous papers by the authors and others, the functions
\begin{equation}
E := (\rho + i g_{12}) / g_{22} \; , \; \Gamma := \frac{1}{2} \ln(-\rho g_{34})
\end{equation}
were used in place of $\E$ and $\gamma$.  (Recall that $E$, like $\E$, is a
solution of the Ernst equation.)  However, we have found that the employment of
$\E$ and $\gamma$ is more suitable for applying and extending the KC group and
for analyzing the gravitational field in a neighborhood of the axis ($\rho =
0$).

The vacuum field equations expressed in terms of the triad of functions $\rho,
\E, \gamma$ are obtained from those expressed in terms of $\rho, E, \Gamma$
simply by substituting $\E$ for $E$ and $\gamma$ for $\Gamma$.  When these
substitutions are applied to a result\cite{10} previously derived by the
authors for the function $\Gamma$, one obtains the following solution for
$\gamma$ in terms of $\E$ and the initial data functions
$r, \E_{3}, s, \E_{2}$:
\begin{equation}
\gamma(u,v)=\gamma_{3}(u)+\gamma_{2}(v)
-\Re \int_{0}^{u} da \int_{0}^{v} db \left\{
\left[ \frac{\E_{a}(a,b)}{2f(a,b)} \right]
\left[ \frac{\E_{b}(a,b)}{2f(a,b)} \right]^{*}
\right\}
\end{equation}i
where $\gamma_{3}$ and $\gamma_{2}$ are determined up to arbitrary constant by
the equations
\begin{subequations}
\begin{equation}
\dot{\gamma}_{3} = [\ddot{r} - 2(1-r) |\dot{\E}_{3}/2f_{3}|^{2}]/2\dot{r}
\end{equation}
and
\begin{equation}
\dot{\gamma}_{2} = [\ddot{s} + 2(1+s) |\dot{\E}_{2}/2f_{2}|^{2}]/2\dot{s}
\end{equation}
\end{subequations}
which generally hold over the domains $0 < u < u_{0}$ and $0 < v < v_{0}$,
respectively.  If the right sides of the above equations are summable over the
intervals $[0,u]$ and $[0,v]$, respectively, then $\gamma_{3}(u)$ and
$\gamma_{2}(v)$ are defined as the integrals of the right sides of the
equations over these intervals.  Thereupon, $\gamma_{3}(u)$ and $\gamma_{2}(v)$
are continuous throughout $0 \le u < u_{0}$ and $0 \le v < v_{0}$,
respectively, and
\begin{equation}
\gamma(0,0)=0 \; , \; \gamma_{3}(u)=\gamma(u,0) \; , \;
\gamma_{2}(v)=\gamma(0,v) \; .
\end{equation}
With our premises, note that the pair of colliding wave conditions (2) in
Sec.\ VA is equivalent to the statement that $\gamma_{3}(u)$ and
$\gamma_{2}(v)$ are $\bold{C}^{1}$ throughout $0 \le u < u_{0}$ and $0 \le v <
v_{0}$, respectively.

\subsection{The Variables $p,q$ and the Functions $\Izz{\E_{j}}$,
$\Izz{\beta_{j}}$ and $\Izz{\psi_{j}}$}

The symbols `$p$' and `$q$', like `$r$' and `$s$', will have two different but
related uses.  However, the context will always indicate which use is intended.

\begin{description}
\item[Definitions:]
In some equations, $p$ and $q$ will denote independent variables which both
have the range $[0,1]$.  The variables $r$ and $s$ will often be expressed as
functions of $p$ and $q$, respectively, as follows:
\begin{equation}
r(p) := -1+2p^{2} \; , \; s(q)=1-2q^{2} \; .
\end{equation}
In other equations, $p$ and $q$ will denote functions whose domains are
$0 \le u < u_{0}$ and $0 \le v < v_{0}$, respectively, and whose values are
given by
\begin{equation}
p(u) := \sqrt{\frac{1+r(u)}{2}} \; , \; q(v) := \sqrt{\frac{1-s(v)}{2}} \; .
\end{equation}
\item[Definitions:]
Let $\izz{\E_{3}}$ and $\izz{\E_{2}}$ denote those functions whose domains are
$0 \le p < 1$ and $0 \le q < 1$, respectively, such that
\begin{equation}
\izz{\E_{3}}(p) := \bold{\E}_{3}(r(p)) \; , \;
\izz{\E_{2}}(q) := \bold{\E}_{2}(s(q)) \; .
\end{equation}
Also, let $\izz{\beta_{3}}$ and $\izz{\beta_{2}}$ denote those functions whose
domains are the same as those of $\izz{\dot{\E}_{3}}$ and $\izz{\dot{\E}_{2}}$,
respectively, such that
\begin{equation}
\izz{\beta_{j}} := \izz{\dot{\E}_{j}}/2\izz{f_{j}} \mbox{ where }
\izz{f_{j}} := \Re \izz{\E_{j}} \; .
\end{equation}
Note that, for real $\bold{\E} = - \exp(2\bold{\psi})$,
\begin{equation}
\izz{\beta_{j}} = \izz{\dot{\psi}_{j}} \; , \;
\izz{\psi_{3}}(p) := \bold{\psi}_{3}(r(p)) \; , \;
\izz{\psi_{2}}(q) := \bold{\psi}_{2}(s(q)) \; .
\end{equation}
\end{description}

\subsection{The Functions $K_{j}$, $\Izz{K_{j}}$, $u_{j}$ and $\Izz{u_{j}}$}

The functions which we shall define below in Sec.\ VD will be used to prove a
theorem in the next Sec.\ VE.
\begin{description}
\item[Definition:]
Null coordinate transformations $u \rightarrow u'$ and $v \rightarrow v'$ will
be called {\em standard} if they are $\bold{C}^{2}$ and satisfy $du'/du > 0$
and $dv'/dv > 0$ throughout $0 \le u < u_{0}$ and $0 \le v < v_{0}$,
respectively.
\end{description}

The truth of the premises in Sec.\ VA is invariant under standard null
coordinate transformations and so is the truth or falsity, as the case may be,
of each of the colliding wave conditions in Sec.\ VA.  Also, the {\em values}
of the functions $K_{j}$ which are defined below are invariant under standard
null coordinate transformations.

\begin{description}
\item[Definitions:]
Let $K_{3}$ and $K_{2}$ denote those functions whose domains are $0 < u <
u_{0}$ and $0 < v < v_{0}$, and whose values are given by
\begin{equation}
K_{3}(u) := \frac{2[1+r(u)] \exp[2\gamma_{3}(u)]}{\dot{r}(u)} \; , \;
K_{2}(v) := \frac{2[1-s(v)] \exp[2\gamma_{2}(v)]}{-\dot{s}(v)} \; .
\end{equation}
Let $\izz{K_{3}}$ and $\izz{K_{2}}$ denote those functions whose domains are $0
< p < 1$ and $0 < q < 1$, respectively, such that
\begin{equation}
\izz{K_{3}}(p(u)) := K_{3}(u) \; , \;
\izz{K_{2}}(q(v)) := K_{2}(v) \; .
\end{equation}
\item[Theorem:]
If $\gamma_{3}(u)$ and $\gamma_{2}(v)$ are continuous throughout $0 \le u <
u_{0}$ and $0 \le v < v_{0}$, respectively, then
\begin{equation}
\izz{K_{j}}(0) := K_{j}(0) := \lim_{a \rightarrow 0} K_{j}(a) = 0
\end{equation}
exists and (as indicated above) equals zero.  Thereupon, $\izz{K_{j}}(x)$ is
continuous throughout $0 \le x < 1$.
\end{description}

{\em Proof:}  Equations (5.4a) and (5.4b) are expressible in the forms
\begin{subequations}
\begin{equation}
e^{2\gamma_{3}} \frac{d}{du}(\dot{r} e^{-2\gamma_{3}})
= 2(1-r) |\dot{\E}_{3}/2f_{3}|^{2}
\end{equation}
and
\begin{equation}
e^{2\gamma_{2}} \frac{d}{dv}(-\dot{s} e^{-2\gamma_{2}})
= 2(1+r) |\dot{\E}_{2}/2f_{2}|^{2} \; .
\end{equation}
\end{subequations}
Therefore,
$$\dot{r} \exp(-2\gamma_{3}) \mbox{ and } -\dot{s} \exp(-2\gamma_{2})$$
are non-decreasing functions.  The rest of the proof uses the mean value
theorem of the differential calculus and is left to the reader.

It is easy to show that Eqs.\ (5.14a) and (5.14b) are expressible in the form
\begin{equation}
2-[x/\izz{K_{j}}(x)] \izz{\dot{K}_{j}}(x) = (1-x^{2}) |\izz{\beta_{j}}(x)|^{2}
\end{equation}
where we have used Eqs.\ (5.9), (5.11) and (5.12) and where $x=p$ when $j=3$
and $x=q$ when $j=2$.  Equation (5.15) will be used in Sec.\ VE.

{}From our premises in Sec.\ VA and from Eqs.\ (5.4a) and (5.4b),
$\gamma_{3}(u)$
and $\gamma_{2}(v)$ are always $\bold{C}^{1}$ throughout $0 < u < u_{0}$ and $0
< v < v_{0}$, respectively.

\begin{description}
\item[Definitions:]
Suppose $\gamma_{3}(u)$ and $\gamma_{2}(v)$ are $\bold{C}^{1}$ throughout $0
\le u < u_{0}$ and $0 \le v < v_{0}$, respectively.  (This is true if and only
if the colliding wave conditions (2) in Sec.\ VA hold.)  Then let $u_{3}$ and
$u_{2}$ denote those functions whose domains are $0 \le u < u_{0}$ and $0 \le v
< v_{0}$, respectively, and whose values are given by
\begin{equation}
u_{3}(u) := \int_{0}^{u} da \exp[2\gamma_{3}(a)] \; , \;
u_{2}(v) := \int_{0}^{v} db \exp[2\gamma_{2}(b)] \; .
\end{equation}
Let $\izz{u_{3}}$ and $\izz{u_{2}}$ denote those functions whose domains are
$0 \le p < 1$ and $0 \le q < 1$, respectively, such that
\begin{equation}
\izz{u_{3}}(p(u)) := u_{3}(u) \; , \;
\izz{u_{2}}(q(v)) := u_{2}(v) \; .
\end{equation}
Equivalently, as one can see from Eqs.\ (5.7), (5.11), (5.12) and (5.16),
\begin{equation}
\izz{u_{j}}(x) := \int_{0}^{x} dy \; y^{-1} \izz{K_{j}}(y) \; .
\end{equation}
\end{description}

Clearly, $u_{3}$ and $u_{2}$ exist and are $\bold{C}^{2}$ throughout $0 \le u <
u_{0}$ and $0 \le v < v_{0}$, respectively.  The null coordinate
transformations $u \rightarrow u_{3}$ and $v \rightarrow v_{3}$ are standard.
The corresponding transformations for $\gamma_{3}$ and $\gamma_{2}$ are
$\gamma_{3} \rightarrow 0$ and $\gamma_{2} \rightarrow 0$.

\subsection{The Sets $\Sigma(n_{3},n_{2}) \subset \Sigma$}

In the remainder of Sec.\ V, we shall no longer take for granted that null
coordinates $u,v$ exist such that the premises in Sec.\ VA hold and we shall
not grant any colliding wave conditions.  The only initial data functions
which are assumed to be given are $\bold{\E}_{3}(r)$ and $\bold{\E}_{2}(s)$ or,
equivalently, $\izz{\E_{3}}(p)$ and $\izz{\E_{2}}(q)$ which are defined by
Eqs.\ (5.6) and (5.8).

\begin{description}
\item[Premise:]
The functions $\izz{\E_{j}}(x)$ are $\bold{C}^{1}$ throughout $0 \le x < 1$.
\end{description}

Granted the above premise, it has been proven as a special case of a broad
theorem that a function $\izz{\E}$ with domain
\begin{equation}
\dom \izz{\E} := \{ (p,q): 0 < \izz{\rho}(p,q) \le 1 \}
\end{equation}
exists such that
\begin{equation}
\izz{\rho}(p,q) = 1 - p^{2} - q^{2} \; , \;
\izz{\E}(p,0) = \izz{\E}_{3}(p) \; , \;
\izz{\E}(0,q) = \izz{\E}_{2}(q) \; ,
\end{equation}
such that the partial derivatives $\izz{\E_{p}}$, $\izz{\E_{q}}$ and
$\izz{\E_{pq}}$ exist and are continuous throughout $\dom \izz{\E}$ and such
that $\izz{\E}$ is a solution of the Ernst equation throughout
$\dom \izz{\E}$.\cite{29}

\begin{description}
\item[Definition:]
Let real numbers $n_{3}$ and $n_{2}$ be defined by the equation
\begin{equation}
|\izz{\beta_{j}}(0)|^{2} = 2 - (1/n_{j})
\end{equation}
where we recall the definition of $\izz{\beta_{j}}$ by Eq.\ (5.9).  Thus,
$-\infty < n_{j} < 0$ or $\frac{1}{2} \le n_{j} \le \infty$.
\end{description}

At this point, the authors considered a variety of additional premises which
led to broad classes of \CGPWP's that appeared to deserve further study.  The
simplest of these additional premises and the only ones which can be treated
here in a reasonable space is contained in the following theorem.

\begin{description}
\item[Theorem:]
Suppose $\izz{\E_{3}}(p)$ and $\izz{\E_{2}}(q)$ are $\bold{C}^{2}$ throughout
$0 \le p < 1$ and $0 \le q < 1$, respectively, with the possible exception of a
finite number of finite step discontinuities in the second derivatives
$\izz{\ddot{\E}_{3}}(p)$ and $\izz{\ddot{\E}_{2}}(q)$.  Then the following
statements (i) and (ii) are equivalent to each other.
\begin{list}{(\theenumiii)}{\usecounter{enumiii}}
\item
There exist null coordinates $u$ and $v$ such that, if we let
\begin{equation}
\E_{3}(u) := \izz{\E_{3}}(p(u)) \; , \; \E_{2}(v) := \izz{\E_{2}}(q(v)) \; ,
\end{equation}
then $r(u)$, $\E_{3}(u)$ and $s(v)$, $\E_{2}(v)$ satisfy all of the premises
and colliding wave conditions given in Sec.\ VA.
\item
Each of the numbers $n_{3}$ and $n_{2}$ satisfies the condition
\begin{equation}
n_{j} = 1 \mbox{ or } 2 \le n_{j} < \infty \; .
\end{equation}
\end{list}
\end{description}

{\em Proof:}  We first prove that statement (i) implies statement (ii).  The
proof will be given in three parts ($\alpha$), ($\beta$) and ($\gamma$).
\begin{description}
\item[($\alpha$)] From Eq.\ (5.9), $\izz{\beta_{j}}(x)$ is $\bold{C}^{1}$
throughout $0 \le x < 1$ with the possible exception of a finite number of
finite step discontinuities in $\izz{\dot{\beta}_{j}}(x)$.  Therefore, from the
theorem which contains Eq.\ (5.13) and from Eqs.\ (5.15) and (5.21),
\begin{equation}
\izz{K_{j}}(x) = c_{j} x^{(1/n_{j})} e^{S_{j}(x)}
\end{equation}
where $c_{j}$ is a positive constant,
\begin{equation}
S_{j}(x) := \int_{0}^{x} dy \; y^{-1} [2-(1/n_{j})-(1-y^{2})
|\izz{\beta_{j}}(y)|^{2}]
\end{equation}
is $\bold{C}^{1}$ throughout $0 \le x < 1$, and
\begin{equation}
\frac{1}{2} \le n_{j} < \infty \; .
\end{equation}
\item[($\beta$)]
Considering the statements immediately after Eq.\ (5.18), we employ $u_{3}$ and
$u_{2}$ as our null coordinates and obtain the following results from Eqs.\
(5.7), (5.17), (5.18), (5.24) and (5.25):
\begin{equation}
\left. \begin{array}{r} \dot{r}(u_{3}) \\ -\dot{s}(u_{2}) \end{array} \right\}
= 4 c_{j}^{-1} x^{2-(1/n_{j})} e^{-S_{j}(x)}
\end{equation}
and
\begin{equation}
\left. \begin{array}{r} \ddot{r}(u_{3}) \\ -\ddot{s}(u_{2}) \end{array}
\right\}
= 4(1-x^{2}) |\izz{\beta_{j}}(x)|^{2}
\left[ c_{j}^{-1} x^{1-(1/n_{j})} e^{-S_{j}(x)} \right]^{2}
\end{equation}
with the understanding that $x=p(u_{3})$ when $j=3$ and $x=q(u_{2})$ when
$j=2$.  Also,
\begin{subequations}
\begin{equation}
\dot{\E}_{j}(u_{j}) = \izz{\dot{\E_{j}}}(x)
\left[ c_{j}^{-1} x^{1-(1/n_{j})} e^{-S_{j}(x)} \right]
\end{equation}
and
\begin{eqnarray}
\ddot{\E}_{j}(u_{j}) & = & [ c_{j}^{-1} e^{-S_{j}(x)} ]^{2} \times \\
	& & \left\{ \izz{\ddot{\E}_{j}}(x) [ x^{1-(1/n_{j})} ]^{2}
	+ \izz{\dot{\E}_{j}}(x) [(1-x^{2}) | \izz{\beta_{j}}(x) |^{2} - 1]
	x^{1-(2/n_{j})} \right\} . \nonumber
\end{eqnarray}
\end{subequations}
\item[($\gamma$)]
The colliding wave conditions $\dot{r}(0)=\dot{s}(0)=0$ and Eqs.\ (5.9),
(5.21), (5.26) and (5.27) imply
\begin{equation}
\frac{1}{2} < n_{j} < \infty \; , \;
\izz{\beta_{j}}(0) = -\frac{1}{2} \izz{\dot{\E}_{j}}(0) \ne 0
\end{equation}
where we have used the convention $\izz{\E}_{j}(0) = -1$.  So, Eqs.\ (5.28) and
the premise of statement (i) that $r(u)$ and $s(v)$ are $\bold{C}^{2}$
throughout $0 \le u < u_{0}$ and $0 \le v < v_{0}$, respectively, imply that
$$1 \le n_{j} < \infty.$$
Note that the above condition on $n_{j}$ is consistent with Eq.\ (5.29a) and
the premise of statement (i) that $\E_{3}(u)$ and $\E_{2}(v)$ are
$\bold{C}^{2}$ throughout $0 \le u < u_{0}$ and $0 \le v < v_{0}$,
respectively, with the possible exception of a finite number of finite step
discontinuities in the second derivatives.  Equation (5.29b), taken together
with Eq.\ (5.21) and the inequality $\izz{\dot{\E}_{j}}(0) \ne 0$ of (5.30),
further imply that $n_{j}$ must lie in the range (5.23).  We have thus proved
that statement (i) implies statement (ii).
\item[($\delta$)]
We next prove that statement (ii) implies (i).  Let new null coordinates $u$
and $v$ be defined by the equations
\begin{equation}
p(u)=u^{n_{3}} \; , \; q(v) = v^{n_{2}} \mbox{ where $0 \le u < 1$ and
$0 \le v < 1$} .
\end{equation}
\end{description}
Then one uses Eqs.\ (5.7) and (5.21), together with the premise that
$\izz{\E_{j}}(x)$ is $\bold{C}^{2}$ throughout $0 \le x < 1$ (with the possible
exception of a finite number of finite step discontinuities in its second
derivative), to prove that the conditions (5.23) imply that $r(u)$, $\E_{3}(u)$
and $s(v)$, $\E_{2}(v)$ satisfy all of the premises and colliding wave
conditions given in Sec.\ VA.  We leave details to the reader.
\begin{description}
\item[Definition]
For any given numbers $n_{3}$ and $n_{2}$ such that $-\infty < n_{j} < 0$ or
$\frac{1}{2} \le n_{j} \le \infty$ ($j=3,2$), let $\Sigma(n_{3},n_{2})$ denote
the set of all members of $\Sigma_{\e}$ for which $\izz{\E_{3}}(p)$ and
$\izz{\E_{2}}(q)$ are $\bold{C}^{2}$ [with the possible exception of a finite
number of finite step discontinuities in $\izz{\ddot{\E}_{3}}(p)$ and
$\izz{\ddot{\E}_{2}}(q)$] throughout $0 \le p < 1$ and $0 \le 1 < 1$,
respectively, and for which $|\izz{\beta_{j}}(0)|^{2} =
|\izz{\dot{\E}_{j}}(0)/2|^{2} = 2-(1/n_{j})$.
\item[Corollary:]
Suppose $\bold{\E} \in \Sigma(n_{3},n_{2})$.  Then the preceding theorem is
equivalent to the statement that $\bold{\E} \in \Sigma_{\e}^{\CW}$ if and
only if $n_{3}$ and $n_{2}$ both lie in the range $\{ 1 \} \cup [2,\infty[$.
Furthermore, if $\bold{\E} \in \Sigma_{\e}^{\CW}$ and null coordinates $u$
and $v$ are defined by Eqs.\ (5.31), then the transformations $u_{3}
\rightarrow u$ and $u_{2} \rightarrow v$ are standard.
\item[Corollary:]
Suppose $\bold{\E} \in \Sigma(n_{3},n_{2})$ and $\bold{\E} \in
\Sigma_{\e}^{\CW}$.  Then the plane wave front at $u=0$ is impulsive if and
only if $n_{3}=1$, and the plane wave front at $v=0$ is impulsive if and
only if $n_{2}=1$.  Furthermore, a shock front occurs at $u=0$ if and only if
\begin{equation}
\mbox{both $n_{3}=1$ and $\izz{\ddot{\E}_{3}}(0) \ne 0$, or $n_{3}=2$.}
\end{equation}
Likewise, a shock front occurs at $v=0$ if and only if both $n_{2}=1$ and
$\izz{\ddot{\E}_{2}}(0) \ne 0$, or $n_{2}=2$.
\end{description}

{\em Proof:}  From Eqs.\ (5.31),
\begin{equation}
\dot{\E}_{3}(u) = \izz{\dot{\E}_{3}}(p(u)) n_{3} u^{n_{3}-1}
\end{equation}
and
\begin{equation}
\ddot{\E}_{3}(u) = \izz{\ddot{\E}_{3}}(p(u)) [n_{3} u^{n_{3}-1}]^{2}
+ \izz{\dot{\E}_{3}}(p(u)) n_{3} (n_{3}-1) u^{n_{3}-2} ,
\end{equation}
where we note from (5.30) that $\izz{\dot{\E}_{3}}(0) \ne 0$.  The rest of the
proof follows from the well known facts that $\dot{\E}(0) \ne 0$ is necessary
and sufficient for an impulsive wave front at $u=0$, whereas $\ddot{\E}_{3}(0)
\ne 0$ is necessary and sufficient for a shock front at $u=0$.  The proof for
the wave front at $v=0$ is similar.

\subsection{The Spectral Functions $\G_{j}(\sigma)$ and $\lambda(\sigma,z_{0})$
for the Real Members of $\Sigma(n_{3},n_{2})$}

We now specialize to real $\bold{\E} = - \exp(2\bold{\psi})$, whereupon Eqs.\
(5.10) hold.  It is convenient to let
\begin{equation}
p_{\sigma} := \sqrt{\frac{1+\sigma}{2}} \; , \;
q_{\sigma} := \sqrt{\frac{1-\sigma}{2}} \; .
\end{equation}
Note that $p_{\sigma}^{2}+q_{\sigma}^{2}=1$ and that the Abel transforms (1.18)
are expressible as
\begin{subequations}
\begin{equation}
\G_{3}(\sigma) = q_{\sigma} \int_{0}^{\pi/2} d\theta \beta_{3}(p_{\sigma} \sin
\theta)
\end{equation}
and
\begin{equation}
\G_{2}(\sigma) = - p_{\sigma} \int_{0}^{\pi/2} d\theta \beta_{2}(q_{\sigma}
\sin
\theta)
\end{equation}
\end{subequations}
where $\izz{\dot{\psi}_{j}}(x) = \izz{\beta_{j}}(x)$ and where we have used
Eqs.\ (1.21).  The inverses of the above transforms are expressible as
\begin{subequations}
\begin{equation}
\izz{\psi_{3}}(p) = \frac{2\sqrt{2}p}{\pi}
\int_{0}^{\pi/2} d\theta \sin \theta g_{3}(-1+2 p^{2} \sin^{2}\theta)
\end{equation}
and
\begin{equation}
\izz{\psi_{2}}(q) = -\frac{2\sqrt{2}q}{\pi}
\int_{0}^{\pi/2} d\theta \sin \theta g_{2}(1-2 q^{2} \sin^{2}\theta) .
\end{equation}
\end{subequations}
Recall the definition of $n_{j}$ by Eq.\ (5.21).  If one grants that
$\izz{\psi_{j}}(x)$ is $\bold{C}^{1}$ throughout $0 \le x < 1$, then
Eqs.\ (5.36a) and (5.36b) imply that $\G_{3}(\sigma)$ is continuous throughout
$-1 \le \sigma < 1$, that $\G_{2}(\sigma)$ is continuous throughout $-1 <
\sigma \le 1$ and that
\begin{equation}
\G_{3}(-1) = \frac{\pi}{2} \izz{\beta_{3}}(0) \; , \;
\G_{2}(1) = -\frac{\pi}{2} \izz{\beta_{2}}(0)
\end{equation}
where
\begin{equation}
|\izz{\beta_{j}}(0)| = \sqrt{2-(1/n_{j})} .
\end{equation}
If $\izz{\psi_{j}}(x)$ is $\bold{C}^{2}$ throughout $0 \le x < 1$, then Eqs.\
(5.35), (5.36a) and (5.36b) imply that $\G_{3}(-1+2 p_{\sigma}^{2})$ is a
$\bold{C}^{1}$ function of $p_{\sigma}$ over the domain $0 \le p_{\sigma} < 1$
and that $\G_{2}(1-2 q_{\sigma}^{2})$ is a $\bold{C}^{1}$ function of
$q_{\sigma}$ over the domain $0 \le q_{\sigma} < 1$.  The following result is
clear.
\begin{description}
\item[Theorem:]
If $\izz{\psi_{3}}(p)$ is $\bold{C}^{2}$ throughout $0 \le p < 1$, then
$\G_{3}(\sigma)$ obeys a H\"{o}lder condition of index $\frac{1}{2}$ on any
given closed subinterval of $-1 \le \sigma < 1$.  If $\izz{\psi_{2}}(q)$ is
$\bold{C}^{2}$ throughout $0 \le q < 1$, then $\G_{2}(\sigma)$ obeys a
H\"{o}lder condition of index $\frac{1}{2}$ on any given closed subinterval of
$-1 < \sigma \le 1$.
\end{description}

Note that Eqs.\ (5.37a) and (5.37b) imply that a sufficient (but not necessary)
condition for $\izz{\psi_{j}}(x)$ to be $\bold{C}^{2}$ throughout $0 \le x < 1$
is that $g_{3}(-1+2 p_{\sigma}^{2})$ and $g_{2}(1-2 q_{\sigma}^{2})$ be
$\bold{C}^{2}$ functions of $p_{\sigma}$ and $q_{\sigma}$ throughout $0 \le
p_{\sigma} < 1$ and $0 \le q_{\sigma} < 1$, respectively.

Let us next consider $\lambda(\sigma,z_{0})$ as given by Eqs.\ (1.22) and
(1.23).  We have, from Eqs.\ (5.38):
\begin{subequations}
\begin{eqnarray}
\lambda(\sigma,z_{0}) & = & - \frac{1}{2} \izz{\beta_{3}}(0)
\ln \left( \frac{z_{0}-\sigma}{1+\sigma} \right)
-\frac{1}{\pi} \int_{-1}^{z_{0}} d\sigma'
\left[ \frac{\G_{3}(\sigma') - \G_{3}(-1)}{\sigma' - \sigma} \right]
\nonumber \\
& & - \frac{1}{\pi} \int_{z_{0}}^{1} d\sigma`
\frac{\G_{2}(\sigma')}{\sigma' - \sigma}
\end{eqnarray}
for $-1 < \sigma < z_{0}$ and
\begin{eqnarray}
\lambda(\sigma,z_{0}) & = & - \frac{1}{2} \izz{\beta_{2}}(0)
\ln \left( \frac{\sigma-z_{0}}{1-\sigma} \right)
-\frac{1}{\pi} \int_{z_{0}}^{1} d\sigma'
\left[ \frac{\G_{2}(\sigma') - \G_{2}(1)}{\sigma' - \sigma} \right]
\nonumber \\
& & - \frac{1}{\pi} \int_{-1}^{z_{0}} d\sigma`
\frac{\G_{3}(\sigma')}{\sigma' - \sigma}
\end{eqnarray}
\end{subequations}
for $z_{0} < \sigma < 1$.  Suppose $\izz{\psi_{3}}(p)$ is $\bold{C}^{2}$
throughout $0 \le p < 1$.  Then, from the preceding theorem and from a theorem
of Plemelj\cite{38}, {\em the first integral on the right side of Eq.\ (5.40a)
defines a function of $\sigma$ which obeys a H\"{o}lder condition of index
$\frac{1}{2}$ on any given closed subinterval of $-1 \le \sigma < z_{0}$.  As
regards the second integral on the right side of Eq.\ (5.40a), it is an
analytic function of $\sigma$ throughout $-1 \le \sigma < z_{0}$}.  Similar
remarks are clearly applicable to the integrals in Eq.\ (5.40b) when
$\izz{\psi_{2}}(q)$ is $\bold{C}^{2}$ throughout $0 \le q < 1$.

The reader can readily pursue further specializations such as granting that
$\izz{\psi_{j}}(x)$ is $\bold{C}^{\infty}$ or even analytic throughout $0 \le x
< 1$.  In any case, except when $\izz{\beta_{3}}(0)=0$ and
$\izz{\beta_{2}}(0)=0$, the function $\lambda(\sigma,z_{0})$ has those
logarithmic singularities at $\sigma=-1$ and $\sigma=1$ which are displayed by
Eqs.\ (5.40a) and (5.40b).  For $\Sigma(n_{3},n_{2}) \subset
\Sigma_{\e}^{\CW}$,
$$|\izz{\beta_{j}}(0)| = 2 \mbox{ or }
\sqrt{3/2} < |\izz{\beta_{j}}(0)| < \sqrt{2} .$$
So there is no escape from the endpoint logarithmic singularities for the class
of \CGPWP's discussed in this section.  In particular, when the axis ($\rho=0$
and $-1 < z < 1$) is accessible, $\bold{\psi}(z,z)=\lambda(z)$ [see Sec.\ IVC]
has logarithmic singularities at $z = \mp 1$.

%
\setcounter{equation}{0}
\section{On the Next Paper}
\subsection{Preliminary}
In the next paper of this series, we shall extend our formalism
to cover the entire Ernst potential set $\Sigma_{\e}$.  The
concepts and methods which were roughly described in Sec.\ I
will be detailed, and we shall attempt to give the reader an
appreciation of how the formalism can be used by means of a few
examples.  At the same time, we shall avoid obscuring the key
ideas by an excess of formal mathematical topics.  In particular,
some difficult existence proofs will be reserved for later papers.

Let us become more specific about the structure of the next paper.
The main new concepts in that paper will be generalizations of
five constructs which occur in the current paper and which will
be reviewed below.

\subsection{The First Construct}
In Eqs.\ (2.13) to (2.17), we introduced the concept of the
spectral potential $\bold{\xi}$ in an arbitrary gauge (not the
special gauge which we chose later).  This concept will be
generalized in our next paper by employing a factorization of
the $\bold{F}$-potential of W.\ Kinnersley and D.\ M.\ Chitre.\cite{1,2,3}
For the case of a real member of $\Sigma_{\e}$, this
factorization is given by Eq.\ (1.1) and the last factor in
Eq.\ (1.1) is $\exp(-\sigma_{3} \bold{\xi})$.  For an
arbitrary member of $\Sigma_{\e}$, the last factor of a
product similar to the one in Eq.\ (1.1) will be a
generalized version of $\exp(-\sigma_{3} \bold{\xi})$ which we
shall define and discuss in our next paper.  The analysis of
this generalized version of $\exp(-\sigma_{3} \bold{\xi})$ will
lead in a natural way to the second construct.

\subsection{The Second Construct}
The starting points of the major deliberations in the current
paper were Eqs.\ (3.9a) and (3.9b) which can be expressed in
the unified form
\begin{equation}
\bold{\xi}(r,s,\sigma^{\pm},z_{0}) = \lambda(\sigma,z_{0})
\mp i \G(r,s,\sigma,z_{0})
\end{equation}
where
\begin{equation}
\G(r,s,\sigma,z_{0}) = \left\{ \begin{array}{ccc}
\G^{(3)}(r,s,\sigma) & \mbox{when} & r < \sigma < z_{0} , \\
\G^{(2)}(r,s,\sigma) & \mbox{when} & z_{0} < \sigma < s ,
\end{array} \right.
\end{equation}
and where $\G^{(3)}$ and $\G^{(2)}$ are defined by the
integrals on the right sides of Eqs.\ (3.9b) and (3.9a),
respectively.\cite{40}  Note that $\G(-1,1,\sigma,z_{0})$ is the
function $\G(\sigma,z_{0})$ which was defined by Eqs.\ (1.18)
and (1.21).  We can obtain a better feeling for the functions
$\G^{(j)}$ by setting $r' = (1-x^{2}) \sigma + x^{2} r$ and
$s' = (1-x^{2}) \sigma + x^{2} s$ in Eqs.\ (3.9b) and (3.9a),
respectively, where $x$ is a new integration variable such
that $0 \le x \le 1$.  Then
\begin{equation}
\G^{(j)}(r,s,\sigma) = - \sqrt{(\sigma-r)(s-\sigma)}
\bar{\bold{\phi}}^{(j)}(r,s,\sigma)
\end{equation}
where, for $-1 < r \le \sigma < s \le 1$ and
$-1 = r < \sigma < s \le 1$,
\begin{subequations}
\begin{equation}
\bar{\bold{\phi}}^{(3)}(r,s,\sigma) = - 2 \int_{0}^{1} dx \;
\bold{\psi}_{r'}((1-x^{2})\sigma+x^{2}r,s)
\end{equation}
and, for $-1 \le 4 < \sigma \le s < 1$ and
$-1 \le r < \sigma < s = 1$,
\begin{equation}
\bar{\bold{\phi}}^{(2)}(r,s,\sigma) = - 2 \int_{0}^{1} dx \;
\bold{\psi}_{s'}(r,(1-x^{2})\sigma+x^{2}s) .
\end{equation}
\end{subequations}
It can be shown from Eq.\ (2.13) that, for $r < \sigma < s$,
the difference
\begin{equation}
\Delta(\sigma) := \G^{(3)}(r,s,\sigma)-\G^{(2)}(r,s,\sigma)
\end{equation}
is independent of $(r,s)$.  Note that $\G^{(3)}(-1,1,\sigma)$
and $\G^{(2)}(-1,1,\sigma)$ are the functions $\G_{3}(\sigma)$
and $\G_{2}(\sigma)$ which were defined by Eqs.\ (1.21).  We
recall from Sec.\ IVC that the vanishing of $\Delta(\sigma)$
is necessary and sufficient for $\bold{\psi}(r,s)$ to have a
continuous extension to the axis points $-1 < r = s < 1$.

When $-1 < r < s < 1$, one can see that the function of
$\sigma$ given by $\G(r,s,\sigma,z_{0})$ has a continuous
extension to the union of $[r,z_{0}[$ with $]z_{0},s]$ and
satisfies
\begin{equation}
\G(\sigma,s,\sigma,z_{0}) = \G(r,\sigma,\sigma,z_{0}) = 0 .
\end{equation}
The second major new construct of the next paper will be a
$2 \times 2$ matrix generalization of $\G$ with the same
domain.  This generalization of $\G$ will be used to
formulate a third construct.

\subsection{The Third Construct}
In Sec.\ IIIB of the current paper, we introduced a simple
Hilbert problem on the real axis interval $[-1,1]$ of the
complex plane.  This is a special case of a simple Hilbert
problem on the real axis interval $[r,s]$ where
$-1 \le r < z_{0} < s \le 1$.  It is convenient at this
stage to restrict the Hilbert problem to the range
$-1 < r < z_{0} < s < 1$.  The values of the Hilbert problem
solutions corresponding to $r=-1$ or to $s=1$ can always be
obtained by computing the limits of the solutions as
$r \rightarrow -1$ or as $s \rightarrow 1$, respectively.

\begin{description}
\item[First Hilbert problem on \mbox{$[r,s]$} when
$-1 < r < z_{0} < s < 1$:]  Suppose that $\bold{\psi}$ is
given.  Then problem is to find $\bold{\xi}(r,s,\tau,z_{0})$
and $\lambda(\sigma,z_{0})$ such that the following two sets
of conditions are satisfied for any fixed $(r,s,z_{0})$ for
which $-1 < r < z_{0} < s < 1$:
\begin{enumerate}
\item The function of $\tau$ given by $\bold{\xi}(r,s,\tau,z_{0})$
is continuous throughout $\bold{D}(r,s,z_{0})$ (the domain defined
by Eq.\ (2.22)), is holomorphic throughout $C-[r,s]$, has the
value $\bold{\psi}(r,s)$ at $\tau = \infty$ and, for any given
$\epsilon > 0$,
$$|\tau-z_{0}|^{\epsilon} \bold{\xi}(r,s,\tau,z_{0})
\rightarrow 0$$
as $\tau \rightarrow z_{0}$ throughout any sequence of values
in $\bold{D}(r,s,z_{0})$.
\item Equation (6.1) holds, where $\G$ is computed from
$\psi$ by using Eqs.\ (3.9a), (3.9b) and (6.2).
\end{enumerate}
\end{description}

The solution of the above Hilbert problem is unique and,
according to the Plemelj theorem, exists.  Thus, the Hilbert
problem defines $\bold{\xi}$ (in our gauge) and $\lambda$
for any given $\bold{\psi}$.  The solution for $\bold{\xi}$
is
\begin{equation}
\bold{\xi}(r,s,\tau,z_{0}) = \bold{\psi}(r,s)
-\frac{1}{\pi} \int_{r}^{s} d\sigma'
\frac{\G(r,s,\sigma',z_{0})}{\sigma'-\tau}
\end{equation}
and $\lambda(\sigma,z_{0})$ is given by the principal value
of the above expression with $\tau$ replaced by $\sigma$.
Alternatively, as can be seen from Eqs.\ (6.1) and (6.6),
$\lambda(\sigma,z_{0})$ is given by Eqs.\ (2.42a) and (2.42b).

The third major new construct of the next paper will be a
$2 \times 2$ matrix generalization of the above Hilbert
problem.  This will be an HHP on $[r,s]$ and its solution
will define $2 \times 2$ matrix generalizations of
$\exp(-\sigma_{3} \bold{\xi})$ and $\exp(\sigma_{3}\lambda)$
corresponding to any given $\bold{\E} \in \Sigma_{\e}$.  The
$2 \times 2$ matrix generalizations of $\exp(-\sigma_{3} \bold{\xi})$
will be used, in turn, to construct (by simple algebraic means)
the special gauge $\Sigma_{F}$ of spectral potentials which were
briefly described in Sec.\ IA.

\subsection{The Fourth and Fifth Constructs}
The fourth construct in the next paper will be the group
$\bar{K}_{(-1,1)}$ of $2 \times 2$ matrix functions $v(\sigma,z_{0})$
which were briefly described in Sec.\ IA.  This group will be
defined algebraically in terms of our generalization of the set
of functions $\exp[\sigma_{3}\lambda(\sigma,z_{0})]$ (though
we caution that $\bar{K}_{(-1,1)}$ is not identical with the
set and that the relation between the matrices $v$ and the
generalization of the functions $\exp(\sigma_{3}\lambda)$ is
of a subtle character).  The group $\bar{K}_{(-1,1)}$ will be
used to formulate the fifth construct which we now introduce.

In Sec.\ IIIE of the current paper, we defined a second Hilbert
problem on $[r,s]$.  In this Hilbert problem, it is $\lambda(\sigma,z_{0})$
which is assumed to be given and the idea is to find
$\bold{\phi} = \bold{\xi}/\mu$.  An alternative version of the
Hilbert problem is given below.  This version is slightly more
complex than the one in Sec.\ IIIE but it is closer to the
generalization in our next paper.

\begin{description}
\item[Second Hilbert problem on \mbox{$[r,s]$} when
$-1 < r < z_{0} < s < 1$:] Let \

$\lambda(\sigma,z_{0})$ be
given.  The problem is to find $\bold{\phi}(r,s,\tau,z_{0})$
and $\bar{\bold{\phi}}(r,s,\sigma,z_{0})$ such that the
following three sets of conditions are satisfied for any
fixed $(r,s,z_{0})$ for which $-1 < r < z_{0} < s < 1$:
\begin{enumerate}
\item The domain of $\bold{\phi}(r,s,\tau,z_{0})$ is
$\bold{D}(r,s,z_{0}) - \{ r,s \}$ but
$$\mu(r,s,\tau)(\tau-z_{0}) \bold{\phi}(r,s,\tau,z_{0})$$
has a continuous extension to the union of $\{ z_{0} \}$ and
$\bold{D}(r,s,z_{0})$.  Furthermore, $\bold{\phi}(r,s,\tau,z_{0})$
is holomorphic throughout $C - [r,s]$ and is zero at $\tau=\infty$.
\item The function of $\sigma$ given by $\bar{\bold{\phi}}(r,s,\sigma,z_{0})$
is continuous throughout $r \le \sigma < z_{0}$ and $z_{0} < \sigma \le s$
and has (at most) a finite step discontinuity at $\sigma=z_{0}$.
\item The equation
\begin{equation}
\bold{\phi}(r,s,\sigma^{\pm},z_{0}) = \bar{\bold{\phi}}(r,s,\sigma,z_{0})
\mp \frac{i \lambda(\sigma,z_{0})}{\sqrt{(\sigma-r)(s-\sigma)}}
\end{equation}
holds for all $\sigma$ in the ranges $r < \sigma < z_{0}$ and
$z_{0} < \sigma < s$.
\end{enumerate}

The solution for $\bold{\phi}$ is given by Eq.\ (3.31) and the
corresponding expression for $\bar{\bold{\phi}}$ is the principal
value of (3.31).

The fifth major new construct of the next paper will be a $2 \times 2$
matrix generalization of the above Hilbert problem and will, in fact,
be the HHP which was briefly described in Sec.\ IA.  The next paper
will also include some simple examples of this HHP corresponding
to various choices of $v \in \bar{K}_{(-1,1)}$.  We shall derive
the solutions of these examples in the next paper.
\end{description}

%
\section*{Acknowledgement}
This work was supported in part by grants PHY-91-16681 and
PHY-92-08241 from the National Science Foundation.

%
\pagebreak

\section*{References}
\begin{enumerate}
\item W.\ Kinnersley and D.\ M.\ Chitre, J.\ Math.\ Phys.\ {\bf 18}, 1538
(1977).
\item W.\ Kinnersley and D.\ M.\ Chitre, J.\ Math.\ Phys.\ {\bf 19}, 1926
(1978).
\item W.\ Kinnersley and D.\ M.\ Chitre, J.\ Math.\ Phys.\ {\bf 19}, 2037
(1978).
\item R.\ Geroch, J.\ Math.\ Phys.\ {\bf 12}, 918 (1971); {\bf 13}, 394
(1972).
\item W.\ Kinnersley, J.\ Math.\ Phys.\ {\bf 14}, 651 (1973).

\item F.\ J.\ Ernst, Phys.\ Rev.\ {\bf 167}, 1175 (1968); {\bf 168}, 1415
(1968); J.\ Math.\ Phys.\ {\bf 15}, 1409 (1974).
\item I.\ Hauser and F.\ J.\ Ernst, J.\ Math.\ Phys.\ {\bf 30}, 2322 (1989).
See Sec.\ IIA.  (In this reference, $\gamma$ denoted the object which we now
denote by $\bold{\psi}-\frac{1}{2}\ln \rho$.  This use of `$\gamma$' was
completely dropped in all of our subsequent papers.  See Ref.\ 10 for another
notational change.)
\item I.\ Hauser and F.\ J.\ Ernst, J.\ Math.\ Phys. {\bf 31}, 871 (1990).
See Sec.\ I.
\item The Ernst potential $\bold{\E}$ is not to be confused with the Ernst
potential $\bold{E}$ which the authors used in some previous papers (e.g.,
Ref.\ 10) and which will be reviewed in Sec.\ VB of this paper.
\item I.\ Hauser and F.\ J.\ Ernst, J.\ Math.\ Phys. {\bf 30}, 872 (1989).
See Sec.\ II.  (In this reference and in Ref.\ 7, $\psi$ denoted the object
which we now denote by $\psi-\frac{1}{2}\ln \rho$.)
\item See Sec.\ VB of Ref.\ 7 for an explanation of why the Ernst potentials
which represent no \CGPWP's are of interest for the study of \CGPWP's.

\item This is a large subgroup of the KC group but it is not the total KC
group.  The total KC group includes elements which transform members of
$\Sigma_{\e}$ into Ernst potentials whose domains are proper subspaces of
$\{ (r,s) : -1 \le r < s \le 1 \}$ (and which are not, therefore, in
$\Sigma_{\e}$).  We shall have more to say about such elements in a
future paper.
\item The $F$-potentials were originally introduced by KC (Ref.\ 2) as
generating functions for their hierarchy of potentials $H^{(n)}$.  They
restricted their spectral parameter $t=(2\tau)^{-1}$ ro real values and
a few of their conventions (e.g., the labeling of the matrix elements)
differ from ours.  Though KC focused on the stationary axisymmetric
electrovac spacetimes, their formalism is readily extended to cover the
case where the commuting Killing vectors which characterize the
spacetime are both spacelike.
\item I.\ Hauser and F.\ J.\ Ernst, J.\ Math.\ Phys.\ {\bf 21}, 1126
(1980).  The $F$-potential and its key properties are here presented
in terms of the notations and conventions of the authors and for both
signatures ($+-$ and $++$) of the $2$-surface of transitivity of the
pair of commuting Killing vectors.  However, in the case of the signature
$++$, the premise in Sec.\ 3 of Ref.\ 14 that the $F$-potential is a
holomorphic function of the non-ignorable coordinates is unnecessarily
restrictive and will not be assumed in the current series of papers.
Moreover, part 3 of the theorem in Sec.\ 3 of Ref.\ 14 is invalid as
stated (and so is its supposed proof in the appendix).  The conclusions
in the theorem are valid only for the simply connected regions which are
defined in Sec.\ 3 of Ref.\ 14.  Multiply connected regions should have
been avoided.
\item F.\ J.\ Ernst, A.\ Garc\'{\i}a-D\'{\i}az and I.\ Hauser, J.\
Math.\ Phys.\ {\bf 29}, 681 (1988).  Section II of this reference
introduces a gauge of $F$-potentials $F(z,\rho,t)$, where $r=z-\rho$,
$s=z+\rho$ and $\tau=(2t)^{-1}$, which is useful for applications of
the KC group for the purpose of generating new \CGPWP\ solutions.  This
gauge is related to our current gauge of $F$-potentials
$\bold{F}(r,s,\tau,z_{0})$ as follows:  $F(z,\rho,t)=\bold{F}(r,s,\tau,0)$
when $-1 \le r < 0 < s \le 1$.  (The statement in Ref.\ 15 that the
$t$-plane points $(2r)^{-1}$ and $(2s)^{-1}$ are branchpoints is not
generally true since, for fixed $z$ and $\rho$, $F(z,\rho,t)$ cannot
always be analytically continued across the real cuts which join
$(2r)^{-1}$ to $(2s)^{-1}$.  However, this statement has no bearing
on the rest of Ref.\ 15.)
\item I.\ Hauser and F.\ J.\ Ernst, J.\ Math.\ Phys.\ {\bf 20}, 362
(1979).  Though this reference deals only with stationary axisymmetric
vacuum spacetimes, the integral equation is easily adapted to the case
when the commuting Killing vectors are both spacelike.
\item See, e.g., Sec.\ III of Ref.\ 15.  The matrices $u(t)$ in Ref.\
15 are related to the matrices $v(\tau)$ in $K_{(-1,1)}$ by
$u(t)=\Delta(t)v(\tau)[\Delta(t)]^{-1}$ where $t=(2\tau)^{-1}$ and
$\Delta(t)=\left( \begin{array}{cc} t & 0 \\ 0 & 1 \end{array} \right)$.
\item D.\ Kramer and G.\ Neugebauer, Commun.\ Math.\ Phys.\ {\bf 10},
132 (1968).
\item G.\ Neugebauer and D.\ Kramer, Ann.\ Phys.\ (Leipzig) {\bf 24},
62 (1969).
\item C.\ M.\ Cosgrove, J.\ Math.\ Phys.\ {\bf 23}, 615 (1982).
Section 4 of this reference explores the conjecture that the
Kramer-Neugebauer involution is a certain analytic continuation of
a subgroup of the KC group.  Section 1 of the reference contains
further points which are pertinent to our objectives.
\item F.\ J.\ Ernst, A.\ Garc\'{\i}a-D\'{\i}az and I.\ Hauser,
J.\ Math.\ Phys.\ {\bf 28}, 2951 (1987).  The Kramer-Neugebauer
involution is reviewed in Sec.\ IV of this reference from the
viewpoint of its application to \CGPWP's.
\item C.\ M.\ Cosgrove, Ph.\ D.\ Thesis, University of Sydney (1979)
(unpublished).
\item C.\ M.\ Cosgrove, in Proceedings of the Second Marcel
Grossmann Meeting on the {\em Recent Developments of General Relativity},
Trieste, Italy, July 5-11, 1979.  This reference and Refs.\ 24 and
25 contain valuable treatments of the B\"{a}cklund transformations
which lie outside (as well as those which lie inside) the KC group.
\item C.\ M.\ Cosgrove, in {\em Gravitational Radiation, Collapsed
Objects, and Exact Solutions}, Proceedings of the Einstein
Centenary Summer School, edited by C.\ Edwards, Lecture Notes in
Physics {\bf 124} (Springer-Verlag, Berlin, 1980) pp.\ 444-453.
\item C.\ M.\ Cosgrove, J.\ Math.\ Phys.\ {\bf 21}, 2417 (1980).
\item D.\ Maison, Phys.\ Rev.\ Lett.\ {\bf 41}, 521 (1978);
J.\ Math.\ Phys.\ {\bf 20}, 871 (1979).
\item G.\ Neugebauer, J.\ Phys.\ A:  Math.\ Gen.\ (letters) {\bf 12},
L67 (1979); {\bf 13}, L19 (1980); J.\ Phys.\ A {\bf 13}, 1737 (1980).
\item I.\ Hauser and F.\ J.\ Ernst, J.\ Math.\ Phys.\ {\bf 31}, 871
(1990).
\item I.\ Hauser and F.\ J.\ Ernst, J.\ Math.\ Phys.\ {\bf 32}, 198
(1991).
\item U.\ Yurtsever, Phys.\ Rev.\ D {\bf 38}, 1706 (1987).
\item U.\ Yurtsever, Phys.\ Rev.\ D {\bf 40}, 329 (1989).
\item S.\ Chandrasekhar and B.\ C.\ Xanthopoulos, Proc.\ R.\ Soc.\
London Ser.\ A {\bf 408}, 175 (1986); {\bf 410}, 311 (1987).
\item C. Hoenselaers and F.\ J.\ Ernst, J.\ Math.\ Phys.\ {\bf 31},
144 (1990).
\item Y.\ Nutku and M.\ Halil, Phys.\ Rev.\ Lett.\ {\bf 39}, 1379
(1977).
\item I.\ Hauser and F.\ J.\ Ernst, J.\ Math.\ Phys.\ {\bf 22},
1051 (1981).  The axis relation is proven in Sec.\ 5 of Ref.\ 35,
where the matrix $u(t)$ is related to $v(\tau)$ as described above
in Ref.\ 17.  (Part of the Geroch conjecture proof in Ref.\ 35 is
invalid since Eqs.\ (61) and (62) in Ref.\ 35 are not equivalent
as claimed.  This defect is remedied in Ref.\ 36.)
\item I.\ Hauser and F.\ J.\ Ernst, ``A new proof of an old
conjecture,'' in {\em Gravitation and Geometry, a Volume in
Honor of Ivor Robinson}, edited by W.\ Rindler and A.\ Trautman
(Bibliopolis, Naples, 1987) pp. 165-214.  The methods of Refs. 35
and 36 will be extended to $\Sigma_{\e}$ (and proofs of the above
assertions concerning $\Sigma_{\e}^{AX}$ and $\Sigma_{\e}^{an}$
can then be provided) in later papers of the current series.
\item See, e.g., Sec.\ II of Ref.\ 10.
\item A good reference for this topic is N.\ I.\ Muskhelishvili,
{\em Singular Integral Equations} (Noordhoff, Groningen, Holland,
1953), Ch.\ 2.  The author's $\bold{\xi}^{\pm}$ is our
$\bold{\xi}(\sigma^{\pm})$.
\item See, e.g., Sec.\ IIA of Ref.\ 7.
\item The integrals in Eqs.\ (3.9a) and (3.9b) are Abel transforms
such as those employed and analysed in Refs.\ 7 and 10.
\end{enumerate}

%
\newpage

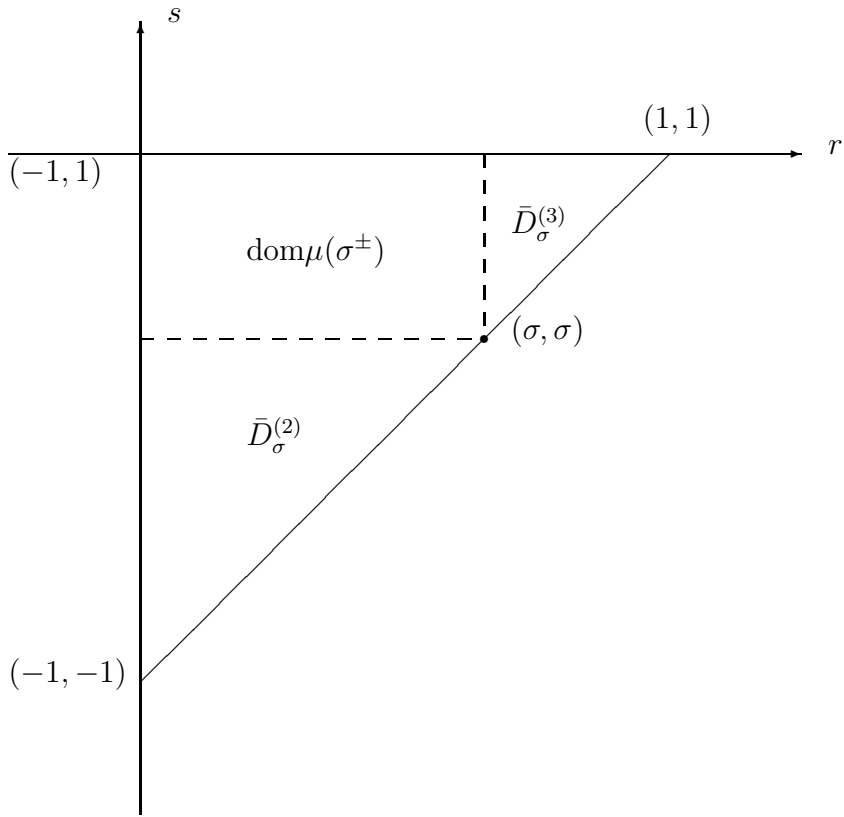
\begin{figure} 
\begin{picture}(350,350)
\put(0,250){\vector(1,0){300}}
\put(50,0){\vector(0,1){300}}
\put(50,50){\line(1,1){200}}
\multiput(50,180)(10,0){13}{\line(1,0){5}}
\multiput(180,250)(0,-10){7}{\line(0,-1){5}}
\put(60,300){$s$}
\put(310,250){$r$}
\put(240,260){$(1,1)$}
\put(0,50){$(-1,-1)$}
\put(190,180){$(\sigma,\sigma)$}
\put(0,240){$(-1,1)$}
\put(90,210){$\dom{\mu(\sigma^{\pm})}$}
\put(90,140){$\bar{D}_{\sigma}^{(2)}$}
\put(190,220){$\bar{D}_{\sigma}^{(3)}$}
\put(180,180){\circle*{3}}
\end{picture}
\vspace{1in}
\caption{The domains of $\mu(\sigma^{\pm})$ and $\mu(\sigma)$ when
$-1 < \sigma < 1$.  The domain of $\mu(\sigma)$ is
$\bar{D}_{\sigma}^{(3)} \cup \bar{D}_{\sigma}^{(2)}$.
The dashed lines separating the domains consist of the points on which
$\mu(r,s,\sigma)=0$.}
\end{figure}

\newpage

\begin{figure} 
\begin{picture}(350,350)
\put(0,250){\vector(1,0){300}}
\put(50,0){\vector(0,1){300}}
\put(50,50){\line(1,1){200}}
\put(50,180){\line(1,0){130}}
\put(180,250){\line(0,-1){70}}
\put(60,300){$s$}
\put(310,250){$r$}
\put(240,260){$(1,1)$}
\put(0,50){$(-1,-1)$}
\put(190,180){$(z_{0},z_{0})$}
\put(0,240){$(-1,1)$}
\put(90,210){$\bold{D}(\tau,z_{0})$}
\put(180,180){\circle*{3}}
\put(170,260){$(z_{0},1)$}
\put(0,180){$(-1,z_{0})$}
\put(50,180){\circle*{3}}
\put(180,250){\circle*{3}}
\end{picture}
\vspace{1in}
\caption{The rectangular domain ${\bf D}(\tau,z_{0})$ when
$\tau \in C-]-1,1[$.  Delete the points $(-1,s)$ when $\tau=-1$.
Delete the points $(r,1)$ when $\tau=1$.}
\end{figure}
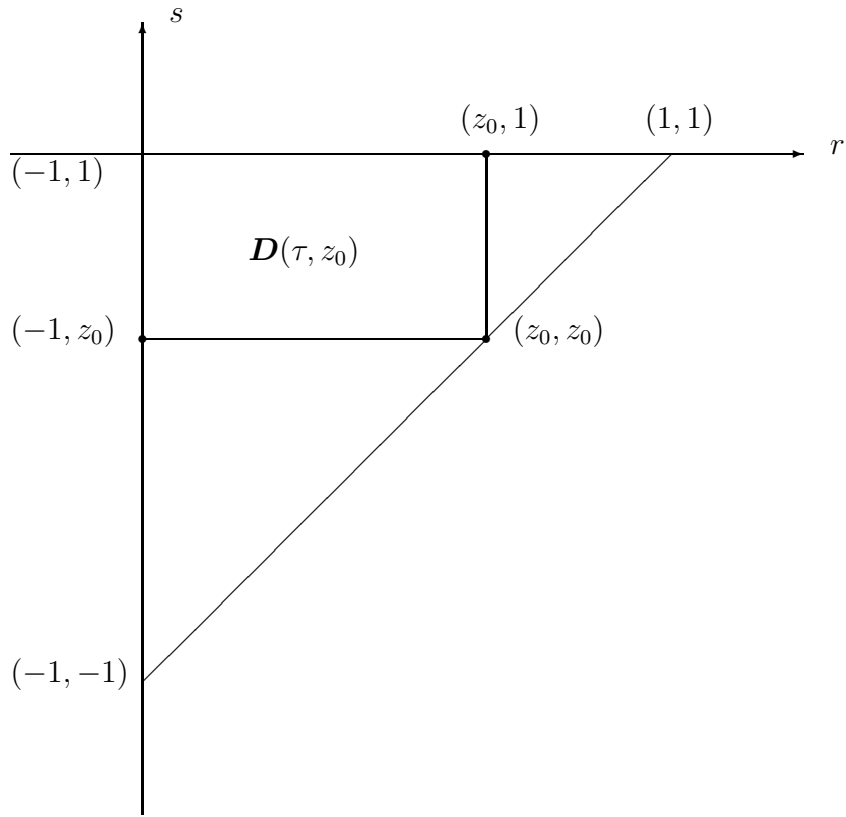

\newpage

\begin{figure} 
\begin{picture}(350,350)
\put(0,250){\vector(1,0){300}}
\put(50,0){\vector(0,1){300}}
\put(50,50){\line(1,1){200}}
\put(50,180){\line(1,0){130}}
\put(180,250){\line(0,-1){70}}
\put(60,300){$s$}
\put(310,250){$r$}
\put(240,260){$(1,1)$}
\put(0,50){$(-1,-1)$}
\put(190,180){$(z_{0},z_{0})$}
\put(0,240){$(-1,1)$}
\put(90,225){$\bold{D}(\sigma^{\pm},z_{0})$}
\put(90,195){$\bold{D}(\sigma,z_{0})$}
\multiput(50,215)(10,0){13}{\line(1,0){5}}
\multiput(180,215)(5,0){7}{\circle*{1}}
\put(225,215){$(\sigma,\sigma)$}
\put(215,215){\circle*{3}}
\put(180,180){\circle*{3}}
\end{picture}
\vspace{1in}
\caption{The rectangular domains ${\bf D}(\sigma^{\pm},z_{0})$
and ${\bf D}(\sigma,z_{0})$ when $z_{0} < \sigma < 1$.
Note that ${\bf D}(\sigma^{\pm},z_{0})$ and
${\bf D}(\sigma,z_{0})$ both contain their common boundary (the dashed
line).}
\end{figure}
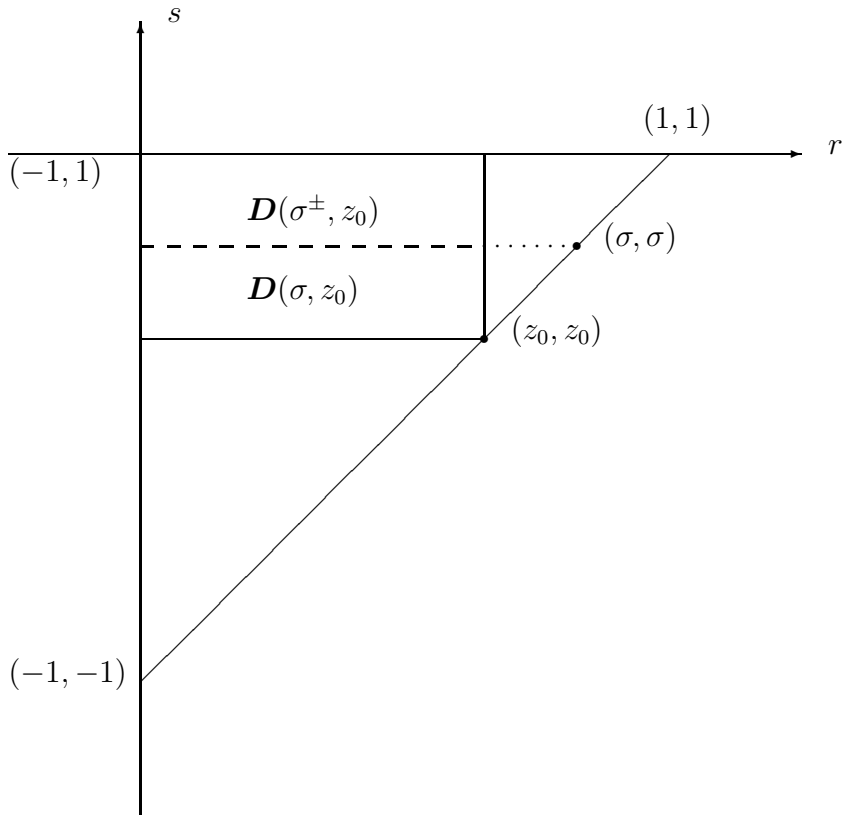

\newpage

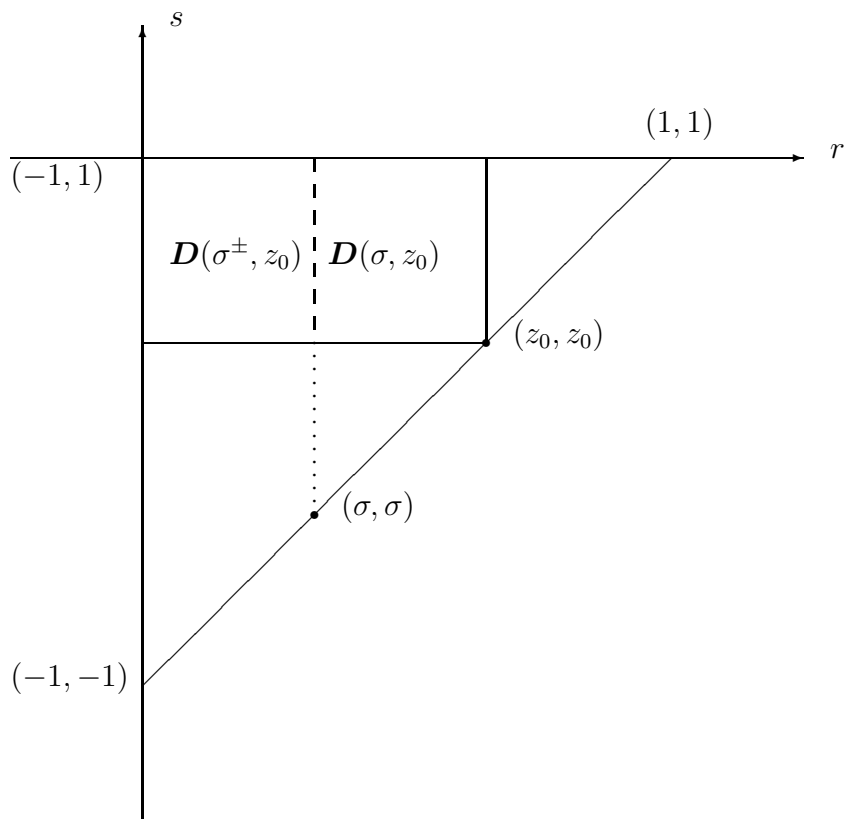
\begin{figure} 
\begin{picture}(350,350)
\put(0,250){\vector(1,0){300}}
\put(50,0){\vector(0,1){300}}
\put(50,50){\line(1,1){200}}
\put(50,180){\line(1,0){130}}
\put(180,250){\line(0,-1){70}}
\put(60,300){$s$}
\put(310,250){$r$}
\put(240,260){$(1,1)$}
\put(0,50){$(-1,-1)$}
\put(190,180){$(z_{0},z_{0})$}
\put(0,240){$(-1,1)$}
\put(60,210){$\bold{D}(\sigma^{\pm},z_{0})$}
\put(120,210){$\bold{D}(\sigma,z_{0})$}
\multiput(115,250)(0,-10){7}{\line(0,-1){5}}
\multiput(115,180)(0,-5){13}{\circle*{1}}
\put(125,115){$(\sigma,\sigma)$}
\put(115,115){\circle*{3}}
\put(180,180){\circle*{3}}
\end{picture}
\vspace{1in}
\caption{The rectangular domains ${\bf D}(\sigma^{\pm},z_{0})$ and
${\bf D}(\sigma,z_{0})$ when $-1 < \sigma < z_{0}$.  Note that ${\bf
D}(\sigma^{\pm},z_{0})$ and ${\bf D}(\sigma,z_{0})$ both contain their
common boundary (the dashed line).}
\end{figure}

\newpage

\begin{figure} 
\begin{picture}(350,350)
\put(0,250){\vector(1,0){300}}
\put(50,0){\vector(0,1){300}}
\put(50,50){\line(1,1){200}}
\multiput(50,180)(10,0){13}{\line(1,0){5}}
\multiput(180,250)(0,-10){7}{\line(0,-1){5}}
\put(60,300){$s$}
\put(310,250){$r$}
\put(240,260){$(1,1)$}
\put(0,50){$(-1,-1)$}
\put(190,180){$(\sigma,\sigma)$}
\put(0,240){$(-1,1)$}
\put(90,210){$\bold{D}(\sigma^{\pm})$}
\put(180,180){\circle*{3}}
\end{picture}
\vspace{1in}
\caption{The domain ${\bf D}(\sigma^{\pm})$ where $-1 < \sigma < 1$.}
\end{figure}
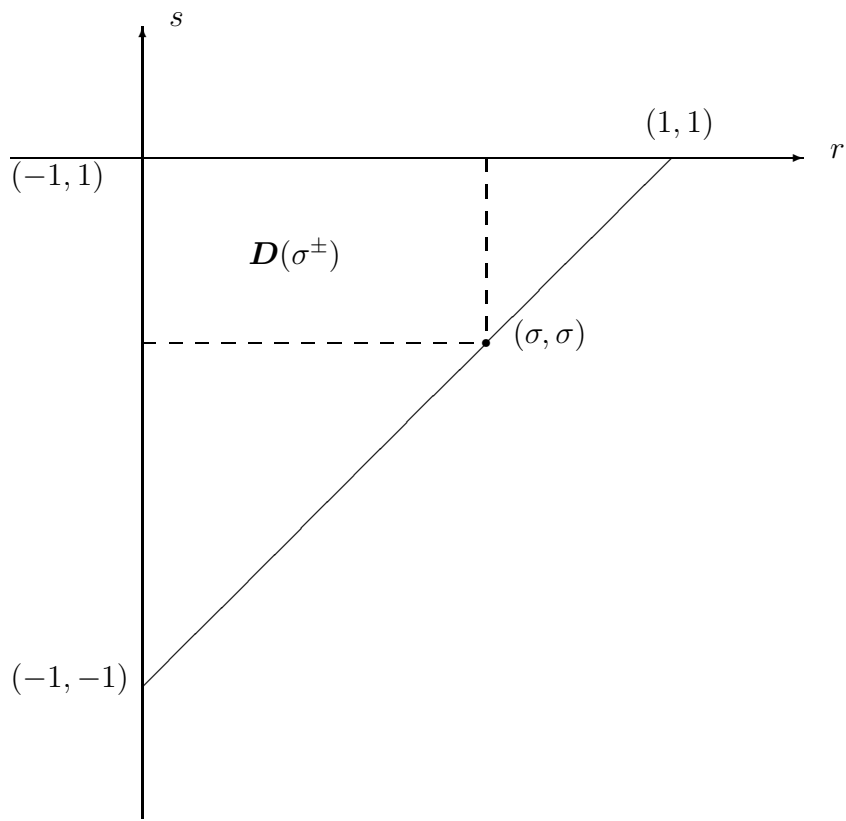

\newpage

\begin{figure} 
\begin{picture}(350,350)
\put(175,175){\oval(200,100)}
\put(75,200){\vector(0,-1){25}}
\put(275,150){\vector(0,1){25}}
\put(175,175){$\Gamma^{+}$}
\put(300,225){$\Gamma^{-}$}
\put(125,220){$\Gamma$}
\end{picture}
\vspace{1in}
\caption{The open sets $\Gamma^{+}$ and $\Gamma^{-}$ which have a given
contour $\Gamma$ as their common boundary such that $\Gamma$,
$\Gamma^{+}$ and $\Gamma^{-}$ are disjoint and have $C$ as their
union.}
\end{figure}
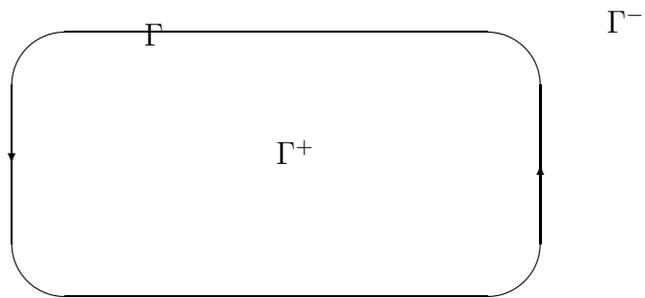

\end{document}